\documentclass[12pt]{iopart}

\usepackage{iopams}
\usepackage[dvipsnames,rgb,dvips]{xcolor}
\usepackage{overpic}
\usepackage{graphicx}
\usepackage{dcolumn}
\usepackage{epstopdf}
\usepackage{siunitx}
\usepackage{mathrsfs}
\makeatletter
\def\amsbb{\use@mathgroup \M@U \symAMSb}
\makeatother
\usepackage{bbm}
\expandafter\let\csname equation*\endcsname\relax
\expandafter\let\csname endequation*\endcsname\relax
\usepackage{amsmath}
\usepackage{soul}

\newcommand{\ve}[1]{\boldsymbol{#1}}
\newcommand{\ma}[1]{\ensuremath{\amsbb{#1}}}

\newcommand\nn{\nonumber}

\newcommand{\tg}{{t=0}}

\pdfoutput=1
\usepackage[dvipsnames,rgb,dvips]{xcolor}
\usepackage{amsmath,amssymb,amsfonts,MnSymbol}
\usepackage{hyperref}
\makeatletter
\def\amsbb{\use@mathgroup \M@U \symAMSb}
\makeatother
\usepackage{mathrsfs}
\usepackage{overpic}

\newcommand{\ee}{\ensuremath{\text{e}}}
\newcommand{\ed}{\ensuremath{\text{d}}}
\newcommand{\dd}[1]{\ensuremath{\tfrac{\text{d}}{\text{d} #1}}}
\newcommand{\ddd}[1]{\ensuremath{\tfrac{\text{d}^2}{\text{d} #1^2}}}
\newcommand{\mc}[1]{\ensuremath{\mathcal{#1}}}
\newcommand{\mbb}[1]{\ensuremath{\mathbb{#1}}}
\newcommand{\eps}{\ensuremath{\varepsilon}}

\newcommand{\sbeqs}[1]{\begin{subequations} #1 \end{subequations}}
\newcommand{\algn}[1]{\begin{align} #1 \end{align}}

\newcommand{\css}[1]{\begin{cases} #1 \end{cases}}
\renewcommand{\tr}{\text{tr}}
\newcommand{\tp}[1]{{#1}^{\sf T}}
\renewcommand{\mc}[1]{\mathscr{#1}}

\newcommand{\eqnlab}[1]{\label{eq:#1}}
\newcommand{\seclab}[1]{\label{sec:#1}}
\newcommand{\applab}[1]{\label{app:#1}}
\newcommand{\figlab}[1]{\label{fig:#1}}
\newcommand{\eqnref}[1]{\eqref{eq:#1}}
\newcommand{\Eqnref}[1]{Eq.~\eqref{eq:#1}}
\newcommand{\Eqsref}[1]{Eqs.~\eqref{eq:#1}}
\newcommand{\secref}[1]{\ref{sec:#1}}
\newcommand{\Secref}[1]{Section~\ref{sec:#1}}
\newcommand{\Secsref}[1]{Sections~\ref{sec:#1}}
\newcommand{\Appref}[1]{\ref{app:#1}}
\newcommand{\figref}[1]{\ref{fig:#1}}
\newcommand{\Figref}[1]{Fig.~\ref{fig:#1}}
\newcommand{\Cite}[1]{Ref.~\cite{#1}}

\renewcommand{\mat}[1]{\begin{pmatrix}#1\end{pmatrix}}

\begin{document}
\title[Fractal catastrophes]{Fractal catastrophes}
\author{J.~Meibohm$^1$, K.~Gustavsson$^1$, J.~Bec$^2$, and B. Mehlig$^1$}
\address{$^1$ Department of Physics, Gothenburg University, 41296 Gothenburg, Sweden}
\address{$^2$ MINES ParisTech, PSL Research University, CNRS, CEMEF, Sophia-Antipolis, France}

\begin{abstract}
We analyse the spatial inhomogeneities (\lq{}spatial clustering\rq{}) in the distribution of particles accelerated by a force that changes randomly in space and time. To quantify spatial clustering, the phase-space dynamics of the particles must be projected to configuration space. Folds of a smooth phase-space manifold give rise to catastrophes (\lq{}caustics\rq{}) in this projection.  When the inertial particle dynamics is damped by friction, however, the phase-space manifold converges towards a fractal attractor. It is believed that caustics increase spatial clustering also in this case, but a quantitative theory is missing. We solve this problem by determining how projection affects the distribution of finite-time Lyapunov exponents. Applying our method in one spatial dimension we find that caustics arising from the projection of a dynamical fractal attractor (\lq{}fractal catastrophes\rq{}) make a distinct and universal contribution to the distribution of spatial finite-time Lyapunov exponents. Our results explain a projection formula for the spatial fractal correlation dimension, and how a fluctuation relation for the distribution of finite-time Lyapunov exponents for white-in-time Gaussian force fields breaks upon projection. We explore the implications of our results for heavy particles in turbulence, and for wave propagation in random media.  \end{abstract} 
\maketitle

\section{Introduction}
There are many situations where ensembles of particles are subject to external forces that appear to fluctuate randomly in space and time. Examples are particles in turbulence, such as water droplets in turbulent clouds~\cite{Bodenschatz2010}, dust in the turbulent gas of protoplanetary disks~\cite{Wilkinson_2008,Anders}, or  small particles floating on the free surface of a fluid in motion~\cite{Som93}.
When the particle momenta are damped by friction, the phase-space dynamics is dissipative, leading to spatial clustering in the form of fractal patterns in the particle distribution in configuration space~\cite{Bec03,Meh04,Gus16}.
Spatial clustering has been observed in experiments \cite{Woo05,Saw08,Sal08,War09,Bal10,Mon10,Gib12,Saw12b} and in numerical simulations of particles in turbulence \cite{Wan93,Hog01,Chu05,Pic05,Sim06,Bec06,Bec07,Cal08,Cal08b,Col09,Saw12a,Saw12b}. The phenomenon is of key significance because it brings particles close together and thus affects the rate of collisions between particles~\cite{Rea00,And07}, their evaporation or condensation~\cite{Bec14}, or chemical reactions~\cite{Kru17}.

The fractal nature of spatial clustering is quantified by fractal dimensions \cite{Ren70,Gra83,Hen83,Bec04,Gra86,Bad87,Gus16} that describe how the fractal patterns fill out configuration space. These dimensions, in turn, are determined by the large-deviation statistics~\cite{Ell07,Hol08,Tou09} of finite-time Lyapunov exponents (FTLEs)~\cite{Gra86,Bad87,Ekdahl}, measuring the evolution of infinitesimal volumes spanned by nearby particle trajectories. The distribution of the FTLEs determines the long-time statistical properties of the dynamically evolving fractal attractor to which the particles converge. 

In the overdamped limit, particle momenta are negligible, so that the phase-space dynamics contracts to configuration space. In this case, the statistical properties of spatial FTLEs is well understood \cite{Bal99,Fal01,Bal01,Bec04}. Inertial particle dynamics, however, occurs in phase space, and the statistical properties of the phase-space attractor are determined by the phase-space FTLEs. To describe {\em spatial} clustering, phase-space volumes must be projected to configuration space. Since it is not understood how this projection affects the distribution of FTLEs, there is no first-principles theory of spatial clustering. The source of the difficulties is well known \cite{Fal02,Wil03,Wil05,Wil06,Gus11b,Gus16} and we illustrate it in \Figref{illustration}. When the inertial phase-space dynamics generates folds~[\Figref{illustration}(a)], the spatial projection becomes many-to-one, causing infinitesimal neighbourhoods of particles to project to single points in configuration space.
\begin{figure}[t]
\centering
	\begin{overpic}{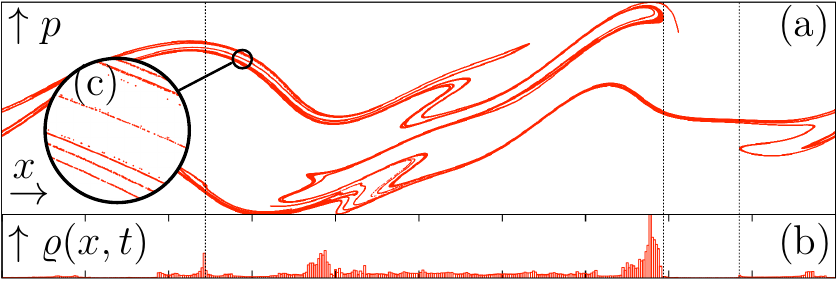}
	\end{overpic}
	\caption{ Fractal clustering in phase-space and configuration-space (schematic). (a) Fractal attractor in phase-spa\-ce (position $x$, momentum $p$) at given time $t$,
	for the one-dimensional statistical model for heavy particles in turbulence reviewed in Ref.~\cite{Gus16}. (b) Increased spatial particle density $\varrho(x, t)$ in the vicinity of caustic folds. (c) The magnification illustrates fractal clustering.
	}\figlab{illustration}
\end{figure}
These singular points are cusp or fold catastrophes, also called \lq{}caustics\rq{} \cite{Wil05,Wil06}, due to their similarities with the random focusing of light in geometrical optics \cite{Ber77,Ber80}. For smooth phase-space manifolds, catastrophes are known to lead to finite-time singularities in the spatial particle density $\varrho(x,t)$~[\Figref{illustration}(b)], suggesting that caustics may increase spatial clustering~\cite{Wil05}. These consideration are, however, too imprecise to quantify spatial clustering. More importantly, these arguments rest on the notion of a smooth manifold, and it is unclear to which extent they apply to fractal phase-space attractors~[\Figref{illustration}(c)]. In other words, it is not understood how caustic folds affect the spatial fractal dimensions, although it is generally assumed that they do. Computing the effect of caustics in the long-time limit is challenging because they give rise to non-perturbative effects \cite{Mei17}, and are therefore thought to cause perturbation expansions for spatial fractal dimensions \cite{Meh04,Dun05,Bec08,Wil10b,Gus15,Gus16,Mei17} to fail.

Here we show how to describe the effect of caustics on spatial clustering from first principles by projecting the phase-space FTLEs to configuration space. We demonstrate our method by deriving the spatial FTLE distribution for inertial particles accelerated by a spatially smooth but random force field in one spatial dimension. Our main result is that caustics give rise to a distinct and universal contribution to the spatial FTLE distribution, independent of the details of the force field.
This caustic contribution results in an exponentially increased probability of observing dense clusters of particles. Furthermore, we demonstrate how caustics affect the distribution of spatial separations, and we show that it explains a projection formula for the spatial fractal correlation dimension. We illustrate the implications of these conceptual insights for white-in-time Gaussian force fields. In this case, there is a fluctuation relation \cite{Che08,Che08b,Sei12,Cel12} that reflects a delicate balance between particle clusters and voids in phase space. We show how this balance is destroyed by the projection, due to additional clusters of particles that stem from caustic catastrophes.

Our results are not confined to dissipative systems, but apply in a limiting case to random dynamical systems without dissipation, such as branched electron flows over a spatially disordered potential \cite{Topinka,Kap02,Met10}, as well as the focusing of light \cite{Ber77,Ber80} and acoustic waves \cite{Whi84,Kul82,Wol00} in random media. In this case fractal clustering is absent, but there are nevertheless substantial spatial inhomogeneities in wave amplitude (and ray location) -- entirely caused by caustics. Our theory predicts the form of the distribution of local stretching factors that determines the spatial patterns formed by the waves \cite{Kap02}. Finally, our results are of interest also in chaos theory, where the distribution of spatial FTLEs is used in the description of deterministic chaotic systems \cite{Gra86,Bad87,Sil07} and in the semiclassical analysis of classically chaotic quantum systems \cite{Sil06,Sil07}. 

The remainder of this paper is organised as follows. \Secref{method} describes the problem, its background, and outlines the methods we use to solve the problem. In \Secref{sp_proj} we explain how to project the FTLEs from phase space to configuration space. The dynamics of phase-space FTLEs in one spatial dimension is derived in \Secref{ps_ftle}. In \Secref{spatialFTLE} we obtain the main results of the paper by applying the projection of the phase-space FTLEs to configuration space in one spatial dimension. In \Secref{fractal} we discuss the consequences of our results for fractal clustering in configuration space and in phase-space. We illustrate the consequences in \Secref{WNL}, by deriving explicit results for the special case of white-in-time Gaussian force fields. In \Secref{discussion} we discuss the physical implications of our results, for particles in turbulence, for the ray dynamics of waves in disordered systems, and for deterministic chaotic systems. Our conclusions are summarised in Section \ref{sec:conclusions}.

\section{Problem formulation and background}\seclab{method}
\subsection{Phase-space finite-time Lyapunov exponents}\seclab{psftle}
Consider the dynamics of the position ${\ve x}_t$ and momentum ${\ve p}_t$ of a particle of mass $m$ in a random force field $\ve f(\ve x,t)$ in $d$ spatial dimensions,
\algn{\eqnlab{eom}
 \dd{t}{\ve x}_t = \frac{\ve p_t}{m}\,,\quad\dd{t}{\ve p}_t= -\gamma \ve p_t+ \ve f(\ve x_t,t)\,.
 }
Here $\gamma$ is a damping coefficient. Equation~\eqnref{eom}  is a widely used model for the dynamics of small, heavy particles in turbulence \cite{Gus16}. In this case the damping is due to viscous friction, and the random force field $\ve f(\ve x,t)$ represents the turbulent fluid-velocity field $\ve u(\ve x,t)$. 

In what follows we use the dimensionless units $t'=t\sqrt{f_0/(m\ell)}$, $\ve x'=\ve x/\ell$, $\ve p'=\ve p/\sqrt{m\ell f_0}$, and $\ve f'=\ve f/f_0$, where $\ell$ and $f_0$ are the correlation length and standard deviation of the force $\ve f$. We drop the primes for notational convenience, and write:
\algn{\eqnlab{eom_dimless}
 \dd{t}{\ve x}_t = {\ve p_t}\,,\quad\dd{t}{\ve p}_t= -\zeta \ve p_t+ \ve f(\ve x_t,t)\,.
 }
Here $\zeta =\gamma\sqrt{m\ell/f_0}$ is a dimensionless damping coefficient. To describe fractal clustering in phase space, we analyse the dynamics of a small neighbourhood of phase-space trajectories around a reference trajectory $({\ve x}_t,{\ve p}_t)$. The phase-space separation between $({\ve x}_t, {\ve p}_t)$ and a neighbouring trajectory $({\ve x}_t',{\ve p}_t')$ is denoted by $\ve R_t=(\delta{\ve x}_t,\delta{\ve p}_t)\equiv({\ve x_t}'-{\ve x_t},{\ve p_t}'-{\ve p_t})$. We consider separations at large times $t$, yet small enough so that the separations are always much smaller than the correlation length of the forcing, $|\delta\ve x_t|\ll1$. Then we can linearise the force field around ${\ve x}_t$ to obtain 
\algn{\eqnlab{resc}
\dd{t}\delta{\ve x}_t = \delta{\ve p}_t\,,\quad\dd{t}\delta{\ve p}_t= -\zeta \delta{\ve p}_t+\mbb{F}({\ve x}_t,t)\delta{\ve x}_t\,,
 }
where $\ma F$ is the random force-gradient matrix with elements $F_{ij} = \partial_j f_i$. The phase-space dynamics in the vicinity of a trajectory $(\ve x_t,\ve p_t)$ becomes
\algn{\eqnlab{flow}
	\dd{t}{\ve R}_t = {\mbb{A}}(t)\,{\ve R}_t\,, \ {\mbb{A}}(t) = \mat{0	&	
    \phantom{-}\ma{I}_{d\times d}	\\	 \mbb{F}(\ve x_t,t)	& -\zeta\,\ma{I}_{d\times d}	&	}\,,
}
where $ \ma{I}_{d\times d}$ is the $d\times d$ unit matrix. The solution of \Eqnref{flow} is expressed in terms of the Green function $\mathbb{J}(t)$ by
\begin{align}
	{\ve R}_t	=\mbb{J}(t)\,{\ve R}_\tg\quad\mbox{with}\quad \mbb{J}(t) = \mc{T}\!\exp \int_0^t \ed t' \mbb{A}(t')\,.
\end{align}
Here $\mc{T}\!\exp$ denotes the time-ordered exponential evaluated along $({\ve x}_t,{\ve p}_t)$. Writing $\mbb{J} = \mbb{V}\,\mbb{R}$, we decompose $\mbb{J}$ into a rotation $\mbb{R}$ and a stretch tensor $\mbb{V}$ with positive and real eigenvalues $(\ee^{\sigma^{(1)}_t t}, \ldots,\ee^{\sigma^{(2d)}_t t})$. The exponents $({\sigma^{(1)}_t}, \ldots,{\sigma^{(2d)}_t})$ are the phase-space FTLEs we intend to calculate. They can be computed as follows. One defines the left Cauchy-Green tensor $\mbb{B} \equiv \mbb{J}\tp{\mbb{J}} = \mbb{V}^2$ with eigenvalues $(\ee^{2\sigma^{(1)}_t t}, \ldots,\ee^{2\sigma^{(2d)}_t t})$. Analysing $\mbb{B}$ instead of $\mbb{V}$ is convenient since $\mbb B$ obeys a closed equation, $\mbb{\dot B} = \mbb{B}\mbb{A}^{\sf T} + \mbb{A} \mbb{B}$, while $\mbb V$ does not. As shown in \Cite{Bal99} this allows to derive evolution equations for the FTLEs and for the orthogonal matrix $\mbb{O}$ that diagonalises $\mbb{B}$. The elements of $\mbb{O}$ can be written as $\mbb{O}_{ij} = [\ve e^{(i)}_t]_j$, where $\ve e^{(i)}_t$ is the eigenvectors of $\ma B$ corresponding to $\sigma^{(i)}_t$. For long enough times, and given a non-degenerate spectrum of Lyapunov exponents, the dynamics of the eigenvectors $\ve e^{(i)}_t$ decouples from that of the FTLEs, leading to a closed set of stochastic equations for $\ve e^{(i)}_t$~\cite{Bal99}:
\algn{\eqnlab{eigenvec}
	\dd{t}\ve e^{(i)}_t = \mbb{A}(t) {\ve e}^{(i)}_t- ({\ve e}^{(i)}_t\cdot \mbb{A}(t){\ve e}^{(i)}_t) {\ve e}^{(i)}_t-\sum_{j<i}^{2d}[{\ve e}^{(j)}_t\cdot (\mbb{A}(t) + \mbb{A}(t)^{\sf T}){\ve e}^{(i)}_t]{\ve e}^{(j)}_t\,.
}
The phase-space FTLEs are obtained as integrals over $\ve e^{(i)}_{t}\cdot \mbb{A}\ve e^{(i)}_{t}$:
\algn{\eqnlab{ftleeqn}
	\sigma^{(i)}_t = \frac1t \int^t_0 \!\!\ed {t'} \,\ve e^{(i)}_{t'}\cdot \mbb{A}(t)\ve e^{(i)}_{t'}\,,
}
It is convenient to arrange the FTLEs in non-increasing order~\cite{Fal01}:
\sbeqs{
	\algn{\eqnlab{constr}
		\sigma^{(1)}_t \ge \sigma^{(2)}_t \ge \cdots \ge \sigma^{(2d)}_t \,.
	}
Since the trace of $\ma A$ is constant, $\tr\,\mbb{A}=-\zeta d$, the phase-space FTLEs obey the sum rule
\algn{\eqnlab{constr_sum}
	\sum_{i=1}^{2d} \sigma^{(i)}_t 	= -\zeta d\,.
	}
}
For ergodic dynamics, the FTLEs have definite limits,
\algn{\eqnlab{psl}
	\lim_{t\to\infty}\sigma^{(i)}_t = \lambda_i\,,
}
the Lyapunov exponents~\cite{Ose68}. In the limit $t\to\infty$, their cumulative sums, $\sum^{n}_{i=1}\lambda_i$, describe the expansion or contraction rates of $n$-dimensional phase-space volumes spanned by $n+1$ nearby particles. The distributions of $\sum_{i=1}^n\sigma^{(i)}_t$, by contrast, describe transient fluctuations of the magnitudes of phase-space (sub-)volumes.
 
\subsection{Large-deviation principle}\seclab{ldp}
At large but finite times $t$, the phase-space FTLEs obey a large-deviation principle~\cite{Ell07,Tou09}. Their joint density has the large-deviation form \cite{Bal99,Bal01,Fal01}
\algn{\eqnlab{ldtFTLE}
	P\left(\sigma^{(1)}_{{t}}=s_1,\ldots,\sigma^{(2d)}_{{t}}=s_{2d}\right) {\propto} 1_{s_1\ge\ldots\ge s_{2d}} \,\delta\left(\sum_{j=1}^{2d} s_j+ \zeta d\right)\, \ee^{-t I(\ve s)}\,,
}
with rate function $I(\ve s)$, and $\ve s =(s_1,\ldots,s_{2d-1})$. The indicator function $1_x$ ensures the ordering of FTLEs, while the Dirac delta function $\delta(x)$ enforces the constraint \Eqnref{constr_sum}. Instead of calculating $I(\ve s)$ directly, it is often easier to compute the scaled cumulant-generating function (SCGF)~\cite{Ell07,Tou09,Che15,Tou18}
\algn{\eqnlab{defLambda}
\Lambda(\ve k) = \lim_{t \to\infty}\frac1t \log\left\langle \exp\left[ t \sum_{i=1}^{2d-1} k_i\sigma^{(i)}_t \right]\right\rangle\,,
}
where $\ve k = (k_1,\ldots,k_{2d-1})$. If $\Lambda(\ve k)$ exists and provided that it is differentiable with respect to $\ve k$, then $I(\ve s)$ is given by the Legendre transform
 ~\cite{Gar77,Ell07},
\algn{\label{eq:ltransform}
	I({\ve s}) = \sup_{\ve k\in\mbb{R}^{2d-1}} \left\{ {\ve k}\cdot {\ve s} - \Lambda(\ve k)	\right\}\,.
}
In \Secref{ps_ftle} we derive stochastic differential equations that allow, in principle, to compute the large-deviation statistics of phase-space FTLEs for a one-dimensional random force field $f(x,t)$. In \Secref{WNL} we show how to solve these equations explicitly, for white-in-time Gaussian force fields.
\subsection{Catastrophes}
Catastrophe theory \cite{Zee79,Arn03,Pos14} is a branch in mathematics that concerns the description and classification of singularities in dynamical systems. The theory explains, for instance, the sensitive parameter dependence of steady-state solutions of differential equations.
Within the theory, singularities arising from folds of a manifold of steady-state solutions over parameter space are categorised into so-called normal forms \cite{Pos14}. An important property of a catastrophe is its codimension, given by the dimension of the space under consideration, minus the dimensionality of the singularity.  Catastrophes of codimension one are called cuspoids (including fold and cusp catastrophes). Although cuspoid catastrophes are the most common ones, catastrophes of higher codimension play an important role in optics \cite{Ber77,Ber80}.
 In optics, catastrophes lead to caustics, singularities in the light intensity due to partial focusing. Caustics arise from the projection of folds of a smooth phase-space manifold onto configuration space. 
 
 \Figref{illustration} illustrates that similar folds, albeit of a \textit{fractal} attractor, are created by the phase-space dynamics \eqnref{eom_dimless}. In \Secref{sp_proj} we show how these fractal catastrophes arise from the spatial projection to configuration space. In \Secref{spatialFTLE} we demonstrate how they affect the distribution of spatial FTLEs.
\subsection{Fractal attractors}\seclab{fractalatt}
When the dynamics is dissipative ($\zeta >0)$, the steady-state phase-space attractor is fractal. This means that the cumulative probability distribution of phase-space separations $R_t =|\ve R_t|=\sqrt{|\delta \ve x_t|^2 + {|\delta \ve p_t|^2}}$ exhibits a power law 
\begin{align}\eqnlab{sepdist}
P(R_t \leq r) \sim r^{D_2}\quad{\mbox{for}\quad r \ll 1}\,.
\end{align}
The exponent $D_2$ defines the phase-space correlation dimension. As can be seen from \Eqnref{sepdist}, the phase-space correlation dimension measures the probability of finding two particles within a distance $r$ in phase-space. For a homogeneous distribution of particles, this probability scales as $\sim r^{2d}$. For fractal particle distributions, on the other hand, the probability scales as $\sim r^{D_2}$ with $D_2<2d$.

The correlation dimension is not the only quantity that measures the fractal properties of particle distributions. Often the Kaplan-Yorke dimension $D_\text{KY}$ \cite{Kap79} is used to characterise the fractal nature of attractors, because $D_\text{KY}$ is defined in terms of the Lyapunov exponents $\lambda_i$. The Kaplan-Yorke dimension $D_\text{KY}$ is thus insensitive to the transient fluctuations determined by $\sigma_t^{(i)}$. For generic non-linear dynamics, $D_\text{KY}$ equals the information dimension $D_1$, but counterexamples can be constructed~\cite{Ott02}.

For the phase-space dynamics \eqnref{eom} that generate the dynamical fractal attractor illustrated in \Figref{illustration}, the distribution of phase-space FTLEs determines not only $D_1$ and $D_2$ but the whole spectrum of fractal phase-space dimensions $D_q$ \cite{Bec04,Gra86,Bad87} for any value of $q$. For the analysis of $D_q$, one considers the moments of the probability ${\mathscr{M}}_r$ contained in a small phase-space ball of radius $r$ around $(\ve x_t,\ve p_t)$. The $r$-scaling of the $n$-th moment of $\mc{M}_r$ measures the probability of finding $n+1$ particles within a distance $r$, thus generalising \Eqnref{sepdist}. The $r$-scaling of $\langle {\mc{M}}_r^n \rangle$ is given by the exponent $\xi_n$ \cite{Bec04,Bec05}\,,
\algn{
	\langle {\mc{M}}_r^n \rangle \sim r^{\xi_n}\quad \text{for} \quad r\ll 1\,.
}
Note in particular that $\langle \mc{M}_r \rangle = P(R_t\leq r)$, so that $\xi_1 = D_2$. More generally, the singularity exponents $\xi_n$ are, by definition, related to the fractal dimensions $D_q$ by $\xi_n = nD_{n+1}$.
 If trajectories do not cross, the local phase-space mass is conserved. In this case  the singularity exponents can be computed from the rate function of FTLEs \cite{Bec04,Gra86,Bad87}. We show in \Secsref{fractal} and \secref{WNL} how to obtain the fractal phase-space dimensions $D_q$ in this way.

The cumulative probability distribution of {\em spatial} separations $|\delta x_t|$ has the form
\begin{align}\eqnlab{spatialdist}
P(|\delta x_t| \leq r) {\propto} r^{\hat D_2}\quad\mbox{for}\quad r\ll 1\,,
\end{align}
where $\hat D_2$ is the spatial correlation dimension \cite{Gus16}\footnote{In what follows, we denote all spatially projected quantities by a hat.}. The spatial correlation dimension $\hat D_2$ measures the probability of finding two particles within a spatial distance $r$. This dimension therefore plays an important role for particle interactions that require spatial proximity.

 More generally, spatial clustering is characterised by the spatial fractal dimensions $\hat D_q$ which describe how the particles fill out configuration space. There is no general formula that connects a spatial fractal dimension $\hat D_q$ to its phase-space counterpart $D_q$. However, it was conjectured on the basis of numerical investigations of spatial clustering~\cite{Bec05,Bec08,Gus16} that the correlation dimension obeys a projection formula of the form
\algn{\eqnlab{projectionformula}
	\hat D_2 = \min \{D_2,d\}\,.
}
For typical projections of generic attractors this relation can be proven to hold for $\hat D_q$ for $0\leq q\leq 2$  \cite{Mar54,Kau68,Hun97}. For $q>2$ one can show that $\hat D_q \leq \min\{D_q,d\}$ \cite{Hun97}. But an important point is that the dynamics \eqnref{eom_dimless} is not isotropic in $2d$-dimensional phase space. Therefore, it is not at all clear whether the projection from phase-space to configuration space is typical, or not. 
 Using our results for the distribution of the spatial FTLE for one spatial dimension, we show in \Secref{fractal} that the spatial correlation dimension $\hat D_2$ obeys the projection formula \eqnref{projectionformula}. More importantly, our theory explains that  $\hat D_2$ saturates at unity (for $d=1$) because of caustics. However, this does not necessarily mean
 that caustics give rise to a spatially uniform distribution of particles, because, possibly, $\hat D_q < 1$ for $q>2$.
\section{Projection to configuration space}\seclab{sp_proj}
In this Section we explain how to project the distribution of phase-space FTLEs to configuration space, taking into account the effect of catastrophes. As mentioned above, the phase-space FTLEs describe how (sub-)volumes evolve in phase space. To understand how they project to configuration space, consider a small $n$-dimensional ($n\leq d$) phase-space volume around a phase-space trajectory $(\ve x_t,\ve p_t)$. Assume that the initial volume lies entirely within configuration space, so that for any vector $\ve w$ in that volume we have $\ve w_{\tg} = \sum_{j=1}^n w_{j} \ve{\hat e}^{(j)}$, where $\ve {\hat e}^{(1)},\ldots,\ve {\hat e}^{(n)}$ is the Cartesian basis in configuration space. At long times the volume aligns with the $n$ eigenvectors 
 $\ve e^{(1)}_t,\ldots,\ve e^{(n)}_t$ of $\ma B$ that correspond to the largest FTLEs, $\sigma_t^{(1)},\ldots,\sigma_t^{(n)}$. As a consequence, the projection of $\ve w_t$ to configuration space evolves as
\algn{
\label{eq:vt}
	\ve {\hat w}_t = \sum_{i,j = 1}^n \ve {\hat e}^{(i)} W^{(n)}_{ij} 
	{\ve e_t^{(j)}\cdot \ve w_{t=0}}\quad \mbox{with}\quad W^{(n)}_{ij} = \ee^{\sigma^{(j)}_t t}\,\ve {\hat e}^{(i)}\cdot\ve e_t^{(j)}\,.
}
The absolute value of the determinant of $\ma W_t^{(n)}$, 
\algn{\eqnlab{detW}
	|\det\mbb{W}_t^{(n)} |= \ee^{t\sum_{i=1}^{n}\sigma^{(i)}_t }|\det \mbb{O}_{t}^{(n)}|\,,
	}
 determines how $n$-dimensional \textit{spatial} volumes expand and contract. Here $\mbb{O}_t^{(n)}$ is the $n\times n$ sub-matrix of $\mbb{O}_t$ corresponding to $\ve e^{(1)}_t,\ldots,\ve e^{(n)}_t$. \Eqnref{detW} shows that $|\det\mbb{W}_t^{(n)} |$ factorises into a phase-space volume factor, $\ee^{t\sum_{i=1}^{n}\sigma^{(i)}_t }$, and a spatial volume factor, $|\det \mbb{O}_t^{(n)}|$. Since $0\leq |\det \mbb{O}_t^{(n)}| \leq 1$, we can write $|\det \mbb{O}_t^{(n)}|=\cos\alpha^{(n)}_t$ and assign periodic boundary conditions to the angle $\alpha^{(n)}_t$ in $[-\pi/2,\pi/2)$. \Eqnref{detW} allows us to express the spatial FTLEs in terms of the phase-space FTLEs as
\algn{\eqnlab{projxkdim}
	\sum_{i=1}^n \hat\sigma^{(i)}_t \equiv \frac1t\log\Bigg|\frac{\det\mbb{W}_t^{(n)}}{\det\mbb{W}_{\tg}^{(n)}}\Bigg| = \sum_{i=1}^n \sigma^{(i)}_t + \frac1t\log \left( \frac{\cos \alpha^{(n)}_t}{\cos \alpha^{(n)}_{\tg}}\right)\,.
}
Using \Eqnref{ftleeqn} we obtain
\sbeqs{\eqnlab{eomgen}
	\algn{
	\sum_{i=1}^n \hat\sigma^{(i)}_t &= \frac1t \int^t_0 \!\!\ed {t'} \,\,\sum_{i=1}^n \ve e^{(i)}_t\cdot\mbb{A}(t) \ve e^{(i)}_t- \frac1t\int_0^t \!\ed \alpha^{(n)}_t \tan \alpha^{(n)}_t\,,\eqnlab{projxkdim2}\\
\dd{t} \alpha_t^{(n)} &= -\left[\tan\alpha_t^{(n)}\right]^{-1}\Tr\left( \dd{t} \mbb{O}^{(n)}_t \Big[	\mbb{O}^{(n)}_t	\Big]^{-1}\right)	\,. \eqnlab{alphaeqn2}
	}
}
Here $\Tr(\ldots)$ denotes the trace of the matrix. Equations~\eqnref{eomgen} describe how folds of the phase-space manifold (catastrophes) affect the spatial FTLEs.
A catastrophe of spatial codimension $d-n+1$ or larger occurs in the spatial subspace spanned by $\ve{\hat e}^{(1)},\ldots,\ve{\hat e}^{(n)}$ when $\cos\alpha^{(n)}_t \to 0$. We denote by $J^{(n)}$ the rate at which $\cos\alpha^{(n)}_t \to 0$, i.e., $\alpha^{(n)}_t$ transitions from $-\pi/2$ to $[\pi/2]^+$. Most importantly, $J \equiv J^{(d)}$ is the rate of formation of catastrophes of codimension one or larger, often called simply \lq{}rate of caustic formation\rq{}~\cite{Wil03,Wil06}. Note that the second integral in \Eqnref{projxkdim2} diverges as $\alpha_t^{(n)}\to-\pi/2$. The time derivative $\dd{t} \alpha_t^{(n)}$, however, remains finite because the factor $\big[\tan\alpha_t^{(n)}\big]^{-1}$ in Eq.~(\ref{eq:alphaeqn2}) cancels the divergence of $\big[	\mbb{O}^{(n)}_t	\big]^{-1}\propto \big[\cos\alpha^{(n)}_t\big]^{-1}$.

In the next Sections we formulate the equations of motion \eqnref{eomgen} explicitly for $d=n=1$. In this case caustics occur at isolated points in configuration space (see \Figref{illustration}).
\section{FTLEs in two-dimensional phase space}\seclab{ps_ftle}
We now apply the methods outlined in the \Secref{method} in one spatial dimension. For $d=1$ there are two eigenvectors $\ve e^{(1)}_t$ and $\ve e^{(2)}_t$ which can be parametrised by a single angle $\alpha_t$:
\algn{\eqnlab{param}
	\ve e^{(1)}_t = \mat{\cos \alpha_t\\\sin \alpha_t}\,,\qquad \ve e^{(2)}_t = \mat{-\sin\alpha_t\\ \cos\alpha_t}\,.
}
The constraint \eqnref{constr_sum} implies that there is only one independent phase-space FTLE, which we take to be $\sigma^{(1)}_t$. In order to derive the dynamics for $\alpha_t$ and $\sigma_t^{(1)}$, we start with \Eqsref{eigenvec} and \eqnref{ftleeqn}:
\algn{
	\dd{t} (t\sigma^{(1)}_t) =\ve e^{(1)}_t\cdot\mbb{A} \ve e^{(1)}_t \quad\mbox{and}\quad \dd{t} \ve e^{(1)}_t = \mbb{A} \ve e^{(1)}_t- (\ve e^{(1)}_t\cdot\mbb{A} \ve e^{(1)}_t)\ve e^{(1)}_t\,.
}
We use the parametrisation \eqnref{param} together with \Eqsref{eigenvec} and \eqnref{ftleeqn} to obtain equations of motion for $\alpha_t$ and $\sigma^{(1)}_t$:
\sbeqs{\eqnlab{1deom}
\algn{
\dd{t} \alpha_t &=-\sin\alpha_t (\sin\alpha_t+\zeta\cos\alpha_t)\!+\! F_t\cos^2\alpha_t\,,\eqnlab{alpha}	\\
\sigma^{(1)}_t & = \frac1t \int_0^t \ed t'\tan\alpha_{t'} + \frac1t \int_0^t\ed\alpha_{t'}\,\tan\alpha_{t'} \,,	\eqnlab{sigma}
}
}
with $F_t\equiv F(x_t,t)$. As explained in the previous Section, caustics occur as $\alpha_t$ transitions from $-\pi/2$ to $[\pi/2]^+$. From \Eqnref{alpha} we see that $\ed\alpha_t|_{\alpha_t = -\pi/2} = -\ed t$, which implies that the transition from $-\pi/2$ to $[\pi/2]^+$ is deterministic with angular velocity $-1$. At the point $\alpha_t=-\pi/2$, both integrands in \Eqnref{sigma} diverge. However, because $\ed\alpha_t|_{\alpha_t = -\pi/2} = -\ed t$ the divergencies in \Eqnref{sigma} cancel, so that the phase-space FTLE $\sigma^{(1)}_t$ remains finite for all times. Equations~\eqnref{1deom} admit the following interpretation, illustrated in \Figref{ellipse}: An initial two-dimensional phase-space disc is deformed by $\ee^{t \sigma^{(1)}_t}$ along $\ve e^{(1)}_t$ and by $\ee^{t\sigma^{(2)}_t}$ along $\ve e^{(2)}_t$. The initial disc is thus squeezed into an ellipse with decreased phase-space volume $\mc{V}_t\sim \mc{V}_{t=0}\ee^{-\zeta t}$, due to the dissipative nature of the dynamics. At the same time, the eigensystem of $\mbb{B}$ rotates by the angle $\alpha_t$.
\begin{figure}[t]
\centering
	\begin{overpic}{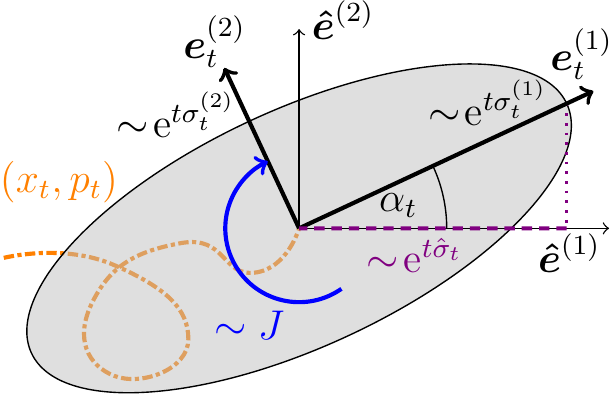}
	\put(0,58.6){\colorbox{white}{\phantom{XX}}}
			\end{overpic}
\caption{
 Evolution of a small, two-dimensional phase-space volume, grey area, around a reference trajectory $(x_t,p_t)$, dash-dotted line. The two
 perpendicular axes are stretched or contracted by $\sim \ee^{t\sigma_t^{(i)}}$, 
 and rotated by $\alpha_t$. The dashed line shows the projection $\sim \ee^{t \hat\sigma_t}$ of the stretching direction to configuration space. The curved arrow indicates the direction of the steady-state flux.}\figlab{ellipse}
\end{figure}

Without imposing strong restrictions on the force gradient $F_t$, we can derive important properties of the probability distribution for $\alpha_t$. If $F_t$ is statistically stationary, $\alpha_t$ reaches a non-equilibrium steady state with density $P_\text{st}(\alpha_t = a)$. As $\alpha_t$ regularly passes $-\pi/2$, $P_\text{st}(\alpha_t = a)$ has a finite flux of magnitude $J$, the rate of caustic formation. The fact that $\ed \alpha_t|_{\alpha_t=-\pi/2} = -\ed t$ then implies
\algn{\eqnlab{pst}
	P_\text{st}\left(\alpha_t=-\pi/2\right) = P_\text{st}\left(\alpha_t= [\pi/2]^+\right)=J\,.
}
In other words, in the presence of caustics, the probability density of $\alpha_t$ at the boundaries of the interval $[-\pi/2,\pi/2)$ is finite and given by the rate of caustic formation.

The system \eqnref{1deom} is more conveniently written in terms of the variable $Z_t = \tan \alpha_t = \delta { p}_t/\delta {x}_t$
\cite{Wil03,Fal02}, which measures the local particle-velocity gradient along the reference trajectory. We obtain
\sbeqs{\eqnlab{1deom2}
\algn{
	\dd{t} Z_t &= -\zeta Z_t -Z_t^2 + F_t\,,	\eqnlab{z} \\
	\sigma^{(1)}_t &= \frac1t \int_0^t \!\!{ \ed t'}\,Z_{t'} + \frac1t \int_0^t \!\! \ed Z_{{t'}}\,\frac{Z_{t'}}{Z_{t'}^2 + 1}\,. \eqnlab{sigma2}
}
}
In terms of the coordinate $Z_t$, a caustic corresponds to $Z_t\to-\infty$ and the immediate re-appearance at $Z_t=+\infty$~\cite{Gus16}. The stochastic dynamics \eqnref{1deom2} determines the distribution of the phase-space FTLE $\sigma^{(1)}_t$, of the form \eqnref{ldtFTLE} with rate function $I(s)$. In one spatial dimension ($d=1$) the large-deviation form of the probability distribution \eqnref{ldp} discussed in \Secref{ldp} reads:
\algn{
	P\left(\sigma^{(1)}_{{t}}=s,\sigma^{(2)}_{{t}}=s'\right) {\propto} \delta(s + s'+\zeta)\theta(s - s')\ee^{-t I(s)}\,.
}
Integrating over $s'$ we obtain the marginal distribution of $\sigma^{(1)}_t$:
\algn{\eqnlab{margftle}
	P\left(\sigma^{(1)}_{{t}}=s\right) {\propto} \theta\left(s+\frac{\zeta}{2}\right)\ee^{-t I(s)}\,.
}
The convex rate function $I(s)$ attains its minimum at $\lambda_1$, where $I(\lambda_1)=0$ so that $\lambda_1 = \lim_{t\to\infty} \sigma^{(1)}_t$ is the maximal Lyapunov exponent, Eq.~(\ref{eq:psl}).

In Section \ref{sec:WNL} we show how to compute $I(s)$ explicitly for white-in-time force fields, using the method of tilted generators~\cite{Tou09,Che15,Tou18}.
\section{Distribution of the spatial FTLE}\seclab{spatialFTLE}
We outlined in \Secref{sp_proj} how to calculate the spatial FTLE $\hat \sigma_t$ by projection. In one spatial dimension this projection is illustrated as the dashed line in \Figref{ellipse}. From \Eqnref{alphaeqn2} we obtain the equation of motion for the projected spatial FTLE $\hat \sigma_t$:
\algn{\eqnlab{projx}
	\hat \sigma_t = \frac1t \int_0^t \ed t' \tan\alpha_{t'} = \frac1t \int_0^t \ed t' Z_{t'} \,.
}
It is easy to see by comparing \Eqsref{projx} and \eqnref{sigma} that the cancellation of the divergencies of the integrals for $\sigma_t^{(1)}$ does not take place for $\hat \sigma_t$. Instead, the spatial FTLE runs into a logarithmic divergence $\hat \sigma_t \sim \log(|t-t_\text{c}|)$ in the vicinity of a caustic at time $t_\text{c}$. 

In the remainder of this Section we present the two main results of this paper, the spatial rate function $\hat I(s)$ and the spatial SCGF $\hat \Lambda(k)$. These quantities describe the transient fluctuations of the spatial FTLE $\hat \sigma_t$.
\subsection{Spatial rate function $\hat I(s)$}\seclab{rf}
\begin{figure}
	\centering	
	\includegraphics{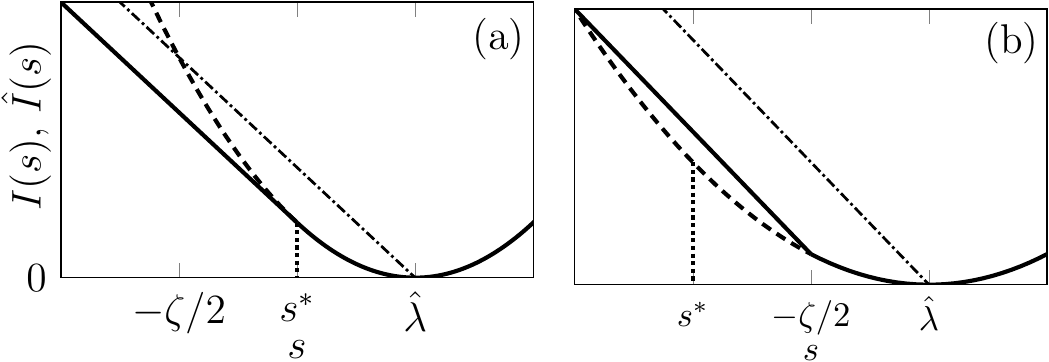}
	\caption{Schematic plot of the phase-space rate function $I(s)$ (dashed line) and the projected configuration-space rate function $\hat I(s)$ (solid line) for different values of $s^*$. The dotted line shows the location of $s^*$. The minimum of $\hat I(s)$ is at the spatial Lyapunov exponent $\hat \lambda$. The dash-dotted line shows the function $\hat \lambda-s$, which is greater than $\hat I(s)$ for $s<\hat \lambda$ (see the discussion in \Secref{discussion}). (a) Case $s^*\geq-\zeta/2$. (b) Case $s^*<-\zeta/2$ where $\hat I(s)$ is not differentiable at $s=0$.}\figlab{ischem}
\end{figure}
We see from \Eqnref{projx} that the spatial projection leads to finite-time divergencies of the spatial FTLE $\hat \sigma_t$. In \Secref{ps_ftle} we have demonstrated that the phase-space FTLE obeys a large-deviation principle with rate function $I(s)$, \Eqnref{ldp}. The question is how the finite-time divergencies of $\hat \sigma_t$ affect the large-deviation principle. In \Appref{hi} we show that the distribution of $\hat \sigma_t$ has indeed a large-deviation form
\algn{
    P({\hat \sigma}_t = s) {\propto} \ee^{-t\hat I(s')}\,,
}
but with an altered, \textit{spatial} rate function $\hat I(s)$. The spatial rate function $\hat I(s)$ is given by
\algn{\eqnlab{hi}
	\hat I(s) = \css{I(s)\,, 	& s \geq \max\{s^*,-\zeta/2\}\,,\\
						-s-\hat I_0\,,		& \text{otherwise}\,,
				}
}
which depends on two constants, $s^*$ and $\hat I_0$. Both these constants depend on the properties of the phase-space rate function $I(s)$. Namely $s^*$ is given by the position of the infimum of $I(s)+s$, 
\algn{\eqnlab{sstar}
s^* = \text{argmin}_{s'\in \mbb{R}}\{I(s') +s'\}\,.
}
while $\hat I_0$ depends on $I(s)$ and the location of $s^*$ relative to $-\zeta/2$:
\algn{\eqnlab{result}
	\hat I_0 = \css{	\Lambda(-1)\,,	&	s^*\geq-\zeta/2\,,	\\
			 \zeta/2- I(-\zeta/2)\,,	& \text{otherwise}\,.}
}
Equation (\ref{eq:hi}) shows that the spatial rate function $\hat I(s)$ is a continuous and convex function of $s$ which coincides with the phase-space rate function $I(s)$ for $s\geq \max\{s^*,-\zeta/2\}$, and is linear otherwise.  However, if $s^*<-\zeta/2$ then $\hat I(s)$ is not differentiable at $s=-\zeta/2$. \Figref{ischem} shows the form of $\hat I(s)$ schematically for these two cases.

The linear part in the spatial rate function $\hat I(s)$ implies that the probability of large, negative values of $\hat\sigma_t$ is exponentially enhanced, by a factor of $\sim \exp(t[I(s) -\hat I(s)])$. The shape of the linear part does not depend upon the details of $F_t$ and is thus a universal contribution due to caustic catastrophes. Put differently, caustics created by fold catastrophes of the phase-space attractor cause additional spatial clustering, with universal properties. 
\subsection{Spatial scaled cumulant-generating function $\hat\Lambda(k)$}\seclab{scgf}
\begin{figure}
	\centering
	\includegraphics{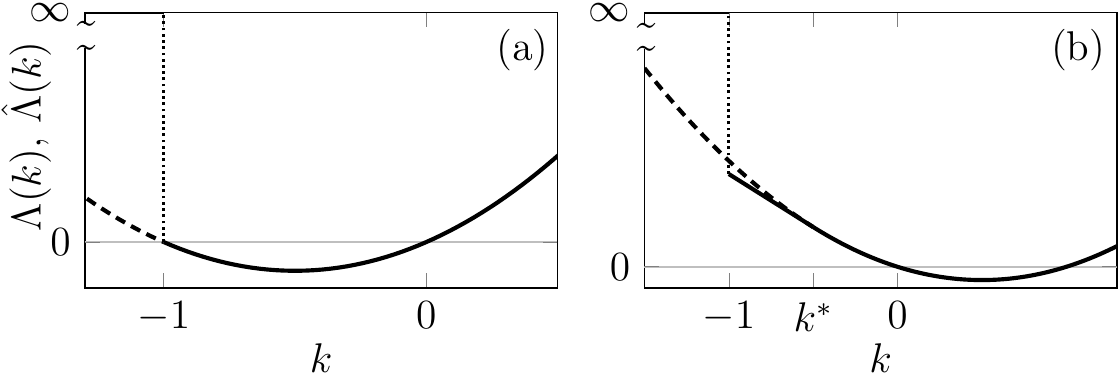}
	\caption{Schematic plot of the phase-space SCGF $\Lambda(k)$ (dashed line) and its spatially projected counterpart $\hat \Lambda(k)$ (solid line) for two different values of $s^*$. (a) Case $s^*\geq-\zeta/2$. (b) Case $s^*<-\zeta/2$ where $\hat \Lambda(k)$ is linear for $k\in(-1,k^*]$.}\figlab{lamschem}
\end{figure}
As the second main result of the paper, we derive the spatial SCGF $\hat \Lambda(k)$. Using \Eqnref{projx} and $Z_t = \dd{t}\log|\delta{x}_t|$, $\hat \Lambda(k)$ can be expressed in terms of the spatial separation $|\delta x_t|$ according to
\algn{\eqnlab{hlkx}
	 \hat \Lambda(k) = \lim_{t\to\infty} \frac1t \log \left\langle \ee^{k \int_0^t \!\!{ \ed t'}\,Z_{t'}}\right\rangle = \lim_{t\to\infty} \frac1t \log \left\langle \left|\frac{\delta{x}_t}{\delta{x}_{\tg}}\right|^k\right\rangle\,.
}
This expression shows that $\hat \Lambda(k)$ is equal to the generalised Lyapunov exponent \cite{Cri88} in one spatial dimension. Since the spatial separations $|\delta x_t|$ contract to zero in finite time, the ensemble average in \Eqnref{hlkx} contains singularities that affect $\hat \Lambda(k)$ for negative $k$. In the next Section we show that $P(|\delta x_t/{\delta} x_{\tg}|=0)>0$ for large enough $t$. This means that $\langle |\delta x_t/\delta x_{\tg}|^{k}\rangle = \int_0^\infty \ed r\, r^{k}P(|\delta x_t/{\delta} x_{\tg}|=r) = \infty$, for $k\leq-1$, which implies $\hat \Lambda(k) = \infty$ for $k\leq-1$. To obtain $\hat\Lambda(k)$ for $k>-1$ we perform a Legendre transform of $\hat I(s)$:
\algn{\eqnlab{leg}
	\hat \Lambda(k) = \sup_{s \in \mbb{R}}\{sk -\hat I(s)\}\,.
}
This calculation is carried out in \Appref{hlam}. We find that the spatial SCGF $\hat \Lambda(k)$ reads
\algn{\eqnlab{hlam}
	\hat \Lambda(k) = \css{			\Lambda(k)\,,	&	k> k^*	\,,\\
								-k\zeta/2- I(-\zeta/2)\,,	&	k^*\geq k> -1	\,,\\
								\infty\,,					& \text{otherwise}\,.
					}
}
The constant $k^*$ is given by
\algn{\eqnlab{kstar}
	k^* = \css{ -1\,, & s^*\geq-\zeta/2\,, \\	I'([-\zeta/2]^+)\,, & \text{otherwise}\,.		}
}
Here $I'([-\zeta/2]^+)$ is the right derivative of $I(s)$ at $s=-\zeta/2$, which enters \Eqnref{kstar} because $I(s)$ is not continuously differentiable at $s=-\zeta/2$ for $s^*<-\zeta/2$.

Equation~\eqnref{hlam} shows that $\hat \Lambda(k)$ coincides with the phase-space SCGF $\Lambda(k)$ for $k>k^*$, has a linear part in the interval $k\in (-1,k^*]$, and diverges for $k\leq-1$. Note that the linear part vanishes when $s^*\geq -\zeta/2$. In \Figref{lamschem} the spatial SCGF is shown schematically for the two cases $s^*\geq-\zeta/2$ and $s^*<-\zeta/2$.

The divergence of $\hat \Lambda(k) $ is the universal analogue of the linear part in the spatial rate function $\hat I(s)$. In view of \Eqnref{hlkx} the divergence of $\hat \Lambda(k)$ for $k\leq-1$ describes a divergence of the negative moments of spatial separations. This divergence is due to a finite value of the probability density of $|\delta x_{t}/\delta x_{\tg}|$ at zero spatial separation $P(|\delta x_{t}/\delta x_{\tg}|=0)>0$, which we derive in the next Section. In other words, the caustic catastrophes allow the particle positions in a neighbourhood to coincide with finite probability.
\section{Relation between phase-space and spatial clustering}\seclab{fractal}
Let us analyse the consequences of \Eqsref{hi} and \eqnref{hlam} for the relation between phase-space and spatial clustering. Recall that the large-deviation statistics of the phase-space FTLEs allows to compute the singularity exponents $\xi_n$, and thus the fractal phase-space dimensions $D_q$. Adapting the formalism described in \Cite{Bec04} to our model, we find that, for a positive maximum Lyapunov exponent $\lambda_1>0$, the exponents $\xi_n$ are given by
\sbeqs{\eqnlab{xi}
\algn{
	\sup_{s\geq-\zeta/2}\{(\xi_n - 2 n)s - {I}(s)\}	&= -\zeta(\xi_n-n)\,,\ \text{if}\ \xi_n > n	\,,\\[-3mm]
	\sup_{s\geq-\zeta/2}\{-\xi_n s - {I}(s)\}	&= 0\,,	\quad\quad\quad\quad \text{otherwise}.
}
}
This result determines the phase-space singularity spectrum $\xi_n$ in terms of the phase-space rate function $I(s)$. Equations~\eqnref{xi} show how the long-time distribution of phase-space FTLEs determines the fractal properties of the phase-space attractor, as discussed in \Secref{fractalatt}.

For spatial clustering, no formula as general as \Eqnref{xi} has been derived, because caustics cause trajectories to cross in configuration space. Therefore, the masses of infinitesimal phase-space volumes are in general not conserved. With the help of \Eqnref{sepdist} one can, however, obtain an analogous formula at least for $D_2 = \xi_1$. It reads \cite{Pik92,noteEkdahl,Gus15}
\algn{
	\sup_{s\in\mbb{R}}\left\{-\hat D_2 s-{\hat I}(s)\right\}=0\,.
}
Comparing with the corresponding condition for the phase-space correlation dimension, \Eqsref{xi} and the projection relation \eqnref{hi} between $I(s)$ and $\hat I(s)$ we deduce that
\algn{\eqnlab{hatD2}
	\hat D_2 = \min\{D_2,1\}\,.
}
This proves that the projection formula \eqnref{projectionformula} holds for $q=2$.
We can also conclude that the saturation of $\hat{D}_2$ at unity is caused by the linear part of the projected rate function, ${\hat I}(s)=-s-\hat I_0$ for $s<s^*$. This means
that the saturation of $\hat D_2$ is a direct consequence of the formation of caustics. 

The behaviour of $\hat D_2$ described by \Eqnref{hatD2} can be explained by comparing the scalings of $P(R_t \leq r)$ and $P(|\delta x_t|\leq r)$ for $r\ll 1$ at large but finite times. In the infinite-time limit, we must then recover \Eqsref{sepdist} and \eqnref{spatialdist}, thereby confirming the relation \eqnref{hatD2} between $D_2$ and $\hat D_2$. We start by considering $P(R_t \leq r)$ for $r\geq\ee^{-\zeta t/2}$ and $D_2 <1$. We write
\algn{\eqnlab{sepdist2}
	P(R_t \leq r) {\propto} \exp\big[-t\inf_{-\zeta/2\leq s\leq t^{-1}\log r} I(s)\big]\,.
}
Since $\lambda_1>0$, typical separations grow exponentially.
In an infinite system the probability of observing any fixed 
value of  $r$ must  tend to zero in the infinite-time limit.  
In order to find the scaling form of the separation distribution that
defines the correlation dimension, we therefore 
demand $\dd{t} P(R_t\leq t^{-1}\log r) = 0$. 
This yields the condition $\sup_{-\zeta/2\leq s < \lambda_1} \{ I'(t^{-1}\log r_0) s - I(s) \} = 0$ in a small range of separations around a small value $r_0$. Comparing with \Eqnref{xi} for $D_2 < 1$ reveals that $I'(t^{-1}\log r_0) = -D_2$. Using this condition we obtain the cumulative separation
distribution by  expanding \Eqnref{sepdist2} around $r_0$. For $|\log (r/r_0)|\ll t$, we find $P(R_t \leq r) \sim r^{-I'(t^{-1} \log r_0)} = r^{D_2}$ for the cumulative distribution of phase-space separations. This shows how the scaling (\ref{eq:sepdist}) of the cumulative distribution
of phase-space separations emerges from our result for the distribution of phase-space FTLEs.

Now consider the cumulative distribution of spatial separations $P(|\delta x_t| \leq r)$. From \Eqnref{hi} we find that for $r\geq \exp(t\max\{s^*,-\zeta/2\})$, $P(|\delta x_t|\leq r) = P(R_t\leq r)$. Differences between the distributions only arise for $r< \exp(t\max\{s^*,-\zeta/2\})$, because this is the regime of the linear part of the spatial rate function $\hat I(s)$. We have in this case $P(|\delta x_t|\leq r) \sim \exp\left[-t \inf_{s \leq t^{-1} \log r}\left\{ -s-\hat I_0\right\}\right]	= r \ee^{\hat I_0 t}$.
As for the phase-space separations, the power-law in $P(|\delta x_t|\leq r)$, \Eqnref{spatialdist}, builds up in a range of separations around a small value $\hat r_0$, given by $\sup_{s \leq \hat \lambda} \{	\hat I'(t^{-1}\log \hat r_0 ) s - \hat I(s)\} = 0$. Comparing this with \Eqnref{hatD2} we conclude that $I'(t^{-1}\log \hat r_0) = - D_2$ for $\hat r_0\geq \exp(t\max\{s^*,-\zeta/2\})$ and $I'(t^{-1}\log \hat r_0) = -1$ otherwise. When $D_2$ transitions from $D_2<1$ to $D_2>1$, $\hat r_0$ moves into the caustic regime with linear $r$-scaling, so that $\hat D_2 = 1$ for $D_2>1$. Hence we recover \Eqnref{spatialdist} in the long-time limit.
For a finite system with boundaries, the cumulative distribution of spatial separations must contain both scalings at small separations, $P(|\delta x_t|\leq r)\approx C_1 r^{D_2} + C_2 r$,  with time-independent constants $C_1$ and $C_2$ \cite{Gus14c}. The small-$r$ scaling of $P(|\delta x_t|\leq r)$ is therefore determined by which of the two powers is dominant. This is consistent with \Eqnref{hatD2}.

In conclusion, caustics affect the behaviour
of  the  distribution of spatial separations $r$ at small values of $r$,
namely that the cumulative distribution depends linearly upon $r$,
$P(|\delta x_t|\leq r)\propto r$ at small separations $r$. This means that
the distribution of separations, $P(|\delta x_t|=r)$, approaches a constant at small separations.
For $\hat D_2<1$, by contrast, this \lq{}caustic regime\rq{} is well separated from a self-similar regime with power-law exponent $D_2$, so that $\hat D_2 = D_2$. For larger values of  $D_2$, the self-similar regime lies within the caustic regime. As a consequence  the $\hat D_2$ must equal  unity. This shows how 
 \Eqnref{hatD2} follows from our results for the distributions of phase-space and spatial FTLEs.
\section{Explicit results for white-in-time Gaussian force fields}\seclab{WNL}
When $f(x,t)$ has Gaussian statistics with zero mean and vanishing correlation time, the gradient $F_t$ is a Gaussian white noise with zero mean and correlation $\langle F_t F_{t'}\rangle = 2 \mc{D}\delta(t-t')$, where $\mc{D}$ is a diffusion constant. In this case, the model depends solely on the dimensionless parameter $\varepsilon^2 =\mc{D}\zeta^{-3}$ and the dynamics of $Z_t$ decouples from that of $(x_t,p_t)$~\cite{Gus16}. This makes it possible to compute the steady-state probability distribution of $\alpha_t$, $P_\text{st}(\alpha_t = a)$, the phase-space SCGF $\Lambda(k)$ and the phase-space rate function $I(s)$ of the phase-space FTLE $\sigma^{(1)}_t$ in explicit form. The explicit results derived in this Section allow us to study in more detail the impact of the spatial projection upon ${\hat I}(s)$, and to evaluate the singularity spectrum $\xi_n$ in \Eqsref{xi}.
\subsection{Steady-state distribution of $\alpha_t$}
\begin{figure}[t]
\centering
	\includegraphics{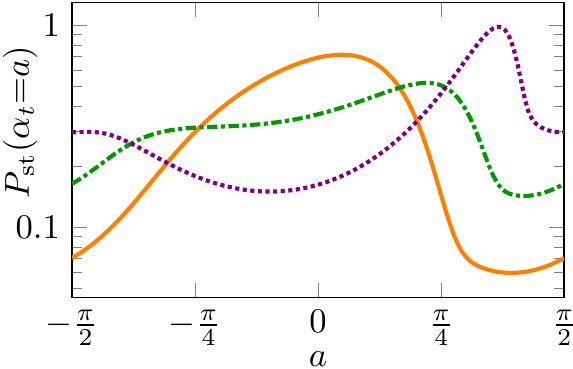}
\caption{Steady-state probability density of $\alpha_t$ with white-in-time force gradients for $\varepsilon=0.5,1,2$ shown as the solid, dash-dotted and dotted line, respectively.}\figlab{palpha}
\end{figure}

Solving the Fokker-Planck equation corresponding to \Eqnref{alpha} (interpreted in the Stratonovich sense) we obtain the steady-state density $P_\text{st}(\alpha_t\!=\! a)$~\cite{Sch02,Wil03,Der07}
\algn{\eqnlab{ness}
P_\text{st}(\alpha_t= a) =\frac{J \zeta^2}{\varepsilon^2}\!\!\int_{-\pi/2}^a \ed a'\, \frac{\ee^{[ U(\tan a')-U(\tan a)]/(\varepsilon^2\zeta^3)}}{(\cos a \cos a')^2}\,,
}
where $U(x) = \zeta x^2/2+x^3/3$. Figure \figref{palpha} shows $P_\text{st}(\alpha_t\!=\! a)$ as a function of $a$ for different values of $\varepsilon$. The rate $J$ of caustic formation is obtained from the normalisation of the probability density in \Eqnref{ness}~\cite{Sch02,Gus16}. For small $\varepsilon$, $\alpha_t$ stays close to zero most of the time, but is more and more likely to approach $-\pi/2$ as $\varepsilon$ increases. It follows from \Eqnref{ness} that $P_\text{st}(\alpha_t\!=\! -\pi/2) = P_\text{st}\left(\alpha_t \!=\! [\pi/2]^+\right) = J$, consistent with \Eqnref{pst}.
\subsection{Explicit calculation of scaled cumulant-generating function}
Since $F_t$ is white in time, $\Lambda(k)$ can be calculated as the leading eigenvalue of a differential operator, the tilted generator $\mc{L}_k$, associated with the large-deviation statistics~\cite{Tou09,Che15,Tou18} of $\sigma_t^{(1)}$. For our case this operator is given by
\algn{\eqnlab{tgen}
\mc{L}_k= (z^2 +1)^{-k/2} \mc{L} (z^2 +1)^{k/2} + k z\,,\quad \mc{L} = \left(-\zeta z -z^2\right)\dd{z} + \eps^2 \zeta^3 \ddd{z}\,,
}
where $\mc{L}$ is the generator of the Markov process \eqnref{z}.

We assume the leading eigenvalue of $\mc{L}_k$ and its adjoint $\mc{L}^\dagger_k$ to be unique and real (we can show this explicitly for even integer $k$, see \Appref{lamk}). In this case, and under certain conditions~\cite{Che15} on the right and left eigenfunctions, $r_k$ and $l_k$, the leading eigenvalue of $\mc{L}_k$ and $\mc{L}^{\dagger}_k$ is given by the phase-space SCGF $\Lambda(k)$, and we have
\algn{\eqnlab{phaseop}
	\mc{L}_k r_k(z) = \Lambda(k) r_k(z) \,,	\quad \mc{L}^{\dagger}_k l_k(z) = \Lambda(k) l_k(z)\,.	
}
Since the dynamics for $\alpha_t$ smoothly transitions from $-\pi/2$ to $[\pi/2]^+$, the corresponding transition for $Z_t = \tan \alpha_t$ from $Z_t \to-\infty$ to $Z_t= \infty$ should also be smooth, so we demand that all eigenfunctions $r_k$ and $l_k$ are symmetric for large $|z|$. 

To find the explicit form of $\Lambda(k)$ we solve \Eqsref{phaseop} numerically by a shooting method -- similar to that described in Refs.~\cite{Gus11b,Gus14c} -- for general $k$. Figure \figref{shooting}(a) shows the resulting $\Lambda(k)$. As expected, the SCGF is convex~\cite{Hol08,Tou09}, yet not a simple parabola as obtained from perturbation theory~\cite{Bal99,Fal01}. For even integers $k$, $\Lambda(k)$ obeys implicit polynomial equations, which we obtain using a method described in \Appref{lamk}. The corresponding exact results for $\Lambda(k)$ are shown as the black dots in \Figref{shooting}(a). We observe perfect agreement with the results obtained from the shooting method. \Figref{shooting}(b) shows the rate function $I(s)$ corresponding to $\Lambda(k)$ obtained by Legendre transform.
\begin{figure}[t]
\centering
	\begin{overpic}{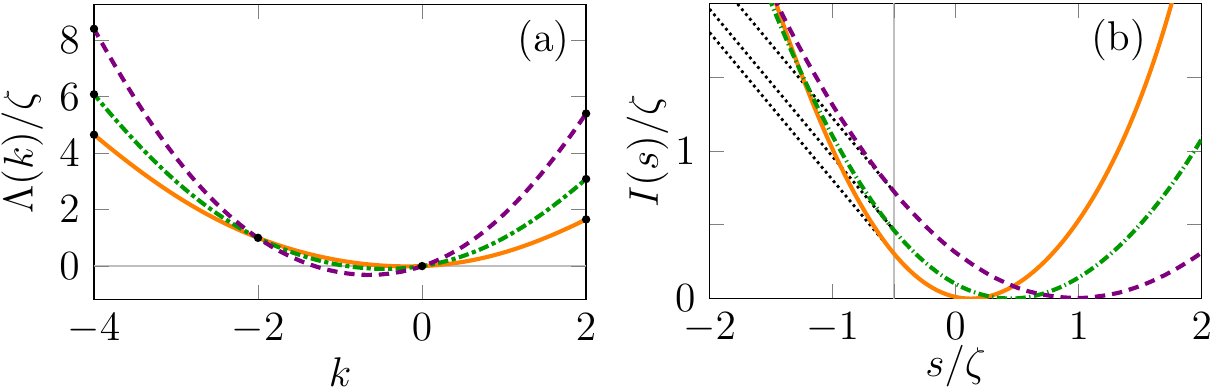}
	\end{overpic}
\caption{Results for white-in-time force gradients with $\varepsilon=2,4,8$ shown as the solid, dash-dotted and dotted line, respectively. (a) SCGF $\Lambda(k)$ obtained from shooting. The dots show solutions of the implicit equations for even integers $k$ (see main text). The light gray line shows $\Lambda(k) = 0$.
 (b) Rate function $I(s)$ obtained from the Legendre transform of $\Lambda(k)$. The dotted lines show the linear part of the corresponding spatial rate function $\hat I(s)$, starting at $s=-\zeta/2$ (light grey line) and extending to $s\to-\infty$.}\figlab{shooting}
\end{figure}
\subsection{Fluctuation relation and spatial rate function}
With help of \eqnref{phaseop},
we can formulate a fluctuation relation \cite{Che08} for $\Lambda(k)$, and carry it over to $I(s)$ using the Legendre transform~(details in \Appref{fluct1}):
\sbeqs{\eqnlab{fluct2}
\algn{
	\Lambda(k-1) - \Lambda(-k-1) &= -\zeta k\,,	\eqnlab{fluct21}\\
	{I}(s-\zeta/2) - {I}(-s-\zeta/2) &= -2 s\,.			\eqnlab{fluct22}
}
}
This relation follows from the time-reversal symmetry of \Eqnref{z},
and it requires that $F_t$ is white in time (see \Appref{fluct2}). 
Fluctuation relations \cite{Che08,Che08b,Sei12} are valuable for characterising fluctuations in non-equilibrium statistical mechanics, because they are some of the few exact and general results that also hold far away from equilibrium. Most of the known relations hold for Markov systems, with few exceptions. The fluctuation relations \eqnref{fluct2} connect the probability of clustering events $(\sigma^{(1)}_t <0)$ to the probability of voids $(\sigma^{(1)}_t >0)$. The inflection point of relation \eqnref{fluct22} at $s=-\zeta/2$ \cite{Fou07} reflects that the system is dissipative, making clusters generally more likely than voids. Equations~\eqnref{fluct2} have interesting consequences for the white-noise limit of our model, which we discuss in the following.

First, \Eqnref{fluct22} allows us to merge the two equations for $\xi_1 = D_2$, \Eqnref{xi} for $n=1$, into one. We find \cite{Bax88,Wil10b,Gus15}
\algn{
	\sup_{s\in\mbb{R}}\{ -D_2 s - I(s) \} = \Lambda(-D_2) = 0\,.
}
Second, \Eqnref{fluct22} allows to determine $s^*$ in \Eqnref{hi} and thus the spatial rate function $\hat I(s)$ for the white-noise case. Differentiating \Eqnref{fluct22} with respect to $s$ and evaluating at $s=0$, we obtain $I'(-\zeta/2) =-1$. Equation~\eqnref{sstar} thus gives $s^* = -\zeta/2$, so that we find from \Eqnref{result}, ${\hat I}_0 = \Lambda(-1)$. Consulting \Eqnref{hi} we conclude that the spatial rate function $\hat I(s)$ in \Eqnref{hi} is linear for $s < -\zeta/2$.  The linear part of the rate function $\hat I(s)$ is shown as the dotted line in \Figref{shooting}(b). 
From \Eqnref{kstar} we obtain $k^*=-1$ so that there is no linear part in $\hat \Lambda(k)$ for white-in-time force gradients. This means that $\hat\Lambda(k) = \Lambda(k)$ for $k>-1$ and $\hat \Lambda(k) = \infty$ for $k\leq -1$.

The linear part of $\hat I(s)$ and the divergence of $\hat\Lambda(k)$ imply that ${\hat I}(s)$ and $\hat\Lambda(k)$ do not obey the fluctuation relations \eqnref{fluct2}. This means that the projection to configuration space destroys the balance described in \Eqsref{fluct2}. The reason is that caustics cause additional clustering, leading to a higher probability of observing particles at small separations, as explained in the previous section. A mathematical interpretation of the broken fluctuation relation is that the spatially projected system loses its Markov property, because the momentum is a hidden variable in the projected space.

We note that the authors of Ref.~\cite{Hub18} computed $\hat \Lambda(k)$ from the leading eigenvalue of an operator $\mc{\hat L}_k$ similar to $\mc{L}_k$, but associated with the spatial FTLE $\hat \sigma_t$, with equation of motion \eqnref{projx}. Our expression for the tilted generator $\mc{L}_k$ in phase space shows that $\mc{L}_k$ and $\mc{\hat L}_k$ have the same leading eigenvalue. This appears to imply that $\Lambda(k) = \hat \Lambda(k)$ for all $k$, at variance with \Eqnref{hlam}. The reason why $\hat \Lambda(k)$ obtained from the $\mc{\hat L}_k$ does not reproduce \Eqnref{hlam} may be that the probabilistic representations of the eigenfunctions of $\mc{\hat L}_k$ are ill-defined due to the finite-time divergence of $\hat \sigma_t$. Therefore, the normalisability requirements for the eigenfunctions of $\mc{\hat L}_k$ given in e.g.~\Cite{Che15} are violated.

\subsection{Fractal phase-space dimensions}
\begin{figure}[t]
\centering
	\begin{overpic}{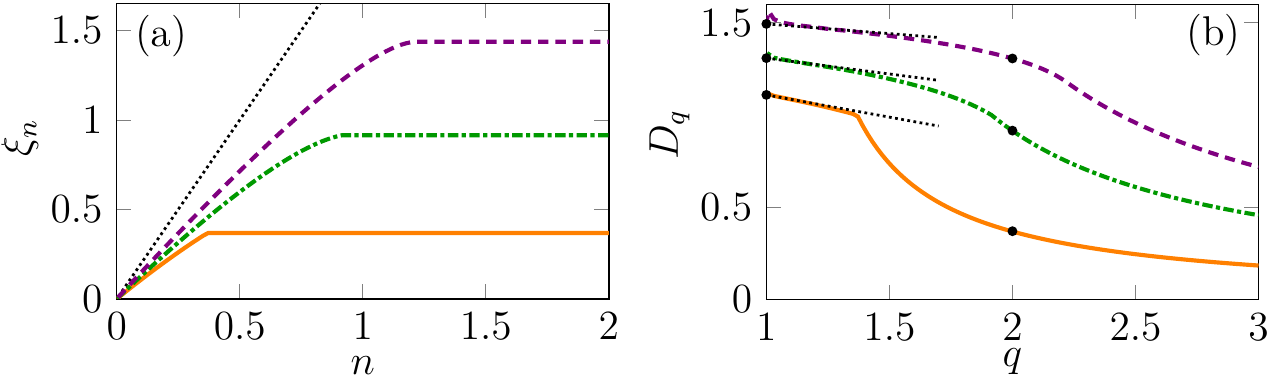}
	\end{overpic}
\caption{Results of white-in-time force gradients for $\varepsilon=2,4,8$ shown as the solid, dash-dotted and dotted line, respectively. (a) Singularity spectrum $\xi_n$ calculated from \Eqnref{xi}. The dashed line shows $\xi_n=2n$ (the singularity spectrum for a homogeneous distribution of particles). (b) Fractal dimensions $D_q$ as a function of $q$. The dots show the values of $D_1$ and $D_2$ obtained by other means. The dotted lines show the asymptotic behaviour around $q=1$ [see \Eqnref{dqasym}].}\figlab{fractal}
\end{figure}
Using \Eqnref{xi} we can now compute the phase-space singularity spectrum $\xi_n$
from our results for $\Lambda(k)$ and $I(s)$ that we obtained from the tilted generator \eqnref{tgen} in the previous Section. Figure \figref{fractal}(a) shows $\xi_n$ for $0\leq n\leq 2$. We observe that $\xi_n$ increases as a function of $n$ and levels off to $\xi_n = \xi_\infty$ for $n>n_\text{crit}$. The behaviour of $\xi_n$ close to $n_\text{crit}$ depends on the value of $\varepsilon$. For small $\varepsilon$ (but larger than $\varepsilon \approx 1.33$, so that $\lambda_1>0$, see \cite{Wil03}) there is a kink at $n_\text{crit}$. In this case, we have $n_\text{crit} = \xi_\infty = \xi_1$. At larger values of $\varepsilon$, on the other hand, $\xi_n$ is smooth around $n_\text{crit}$, and 
$n_\text{crit}$ and $\xi_\infty$ are functions of $\Lambda(-1)$: $n_\text{crit} = 1- \tfrac{2}{\zeta}\Lambda(-1)$ and $\xi_\infty = 1- \tfrac{4}{\zeta}\Lambda(-1)$. These two different behaviours occur below and above a critical value, $\varepsilon_\text{crit}\approx 4.548$, for which $\Lambda(-1)|_{\varepsilon = \varepsilon_\text{crit}} = 0$.

For a homogeneous distribution of particles one has $\langle \mc{M}^n_r \rangle_\text{hom} \sim r^{2n}$. The corresponding singularity exponent, $2n$, is shown as the dotted line in \Figref{fractal}(a). As can be seen in the Figure, $\xi_n<2n$, for $n>0$. Hence, $\langle \mc{M}^n_r \rangle_\text{hom} \ll \langle \mc{M}^{n}_r \rangle$ for $r\ll 1$. This shows that fractal clustering increases the probability of finding particles close together. Furthermore, the singularity exponent $\xi_n$ is a non-linear function of $n$, which implies anomalous scaling of the phase-space mass moments $\langle\mc{M}^n_r\rangle$ in $n$. Since Gaussian-distributed mass moments scale as $\sim r^{\xi_2 n/2}$, with exponent linear in $n$ (similarly to $\langle \mc{M}^n_r \rangle_\text{hom} \sim r^{2n}$), we conclude that the mass distribution has non-Gaussian tails, even though the driving force is Gaussian~\cite{Bec04,Bec05}. 

\Figref{fractal}(b) shows the phase-space fractal dimension for $D_q$ for $1\leq q\leq 3$, obtained from the singularity exponent by $D_q = \xi_{q-1}/(q-1)$. The black dots show the Kaplan-Yorke dimension ~\cite{Kap79} $D_\text{KL} = 1 + \lambda_1/(\lambda_1+\zeta)$ and $D_2$~\cite{Gus16}. For fractal distributions with a single scale (mono-fractals) the fractal dimension is independent of $q$, which is clearly not the case here. Instead the phase-space attractor~[\Figref{illustration}(a)] is multi-fractal, in accordance with the numerical observations in~\Cite{Bec05}. Expanding \Eqnref{xi} around $n=0$ we obtain the asymptotic behaviour of $D_q$ around $q=1$. We find to first order in $q-1$
\algn{\eqnlab{dqasym}
	D_{q} \sim 1+\frac{\lambda_1}{\lambda_1+\zeta} -\frac{\zeta ^2 \Lambda''(0)}{2 (\zeta +\lambda_1)^3}(q-1)\,.
}
To leading order, $D_1$ calculated from \Eqnref{xi} recovers the Kaplan-Yorke dimension, so that $D_1 = D_\text{KL}$ \cite{Bec04}. This is the generic case \cite{Ott02}, as mentioned in \Secref{fractalatt}. The linear order in $q-1$ is determined not only by $\Lambda'(0) = \lambda_1$, but also by the second derivative $\Lambda''(0)$. Similarly, terms of order $m\geq2$ in $(q-1)$ can be shown to contain derivatives of $\Lambda(k)$ at $k=0$ of order up to $m+1$. This shows that the non-Gaussian fluctuations of the phase-space FTLE $\sigma^{(1)}_t$ governed by the higher cumulants play a significant role in determining $D_q$ for $q>1$. In the direct vicinity of $q=1$, on the other hand, non-Gaussian fluctuations are insignificant, and $D_q$ is well described by \Eqnref{dqasym}. In \Cite{Sch02} it was shown that $m$-th order cumulant can be obtained analytically, up to multidimensional integration. 
We use this method to calculate $\Lambda''(0)$. Furthermore, $\lambda_1 = \Lambda'(0)$ is known in closed form \cite{Gus16}. We thus obtain from \Eqnref{dqasym} the asymptotics 
of $D_q$ around $q=1$. The result is shown as the dotted line in \Figref{fractal}(b). We observe good agreement around $q=1$ between the results of the two different methods.

In conclusion, the analysis of the phase-space fractal dimensions shows that the phase-space fractal attractor in  \Figref{illustration} has an intricate structure, even for white-in-time Gaussian random force fields. The fractal dimensions $D_q$ are sensitive to non-Gaussian fluctuations of the phase-space FTLE $\sigma^{(1)}_t$ everywhere, except around $q=1$.
\section{Discussion}\seclab{discussion}
In this Section, we discuss the implications of our findings for a range of physical systems, starting with heavy particles in turbulence.
\subsection{Heavy particles in turbulence}\seclab{aerosols}
Equation~\eqnref{eom_dimless} is a model for small heavy particles in turbulence
subject to viscous friction. For small particles Stokes'  law determines
 the viscous damping parameter $\gamma$. The force field $\ve f(\ve x,t)$
 represents the incompressible turbulent fluid-velocity field  $\ve u (\ve x,t)$. Its 
 correlation length $\ell$ is related to the Kolmogorov length \cite{Gus16}. 
 This model has been used to study spatial clustering and caustic formation for particles
 in turbulence, and their consequences for  collisions in turbulent aerosols \cite{Fal02,Wil03,Meh04,Bec05,Dun05,Wil05,Wil06,Bec08,Wil10b,Gus11b,Gus14c,Pum16,Mei17}. 
 Spatial clustering affects rate of particle collisions through the  radial distribution function $g(r)$ evaluated at the contact distance (equal
 to $2a$ for two particles of radius $a$)  \cite{Sun97}.

  In $d$ spatial dimensions the radial distribution function reads $g(r) = P(\hat R_t = r)/r^{d-1}$ \cite{Bec05,Gus14c,Pum16}, where $\hat R_t = |\delta \ve{x}_t|$ is the spatial separation between the centers of mass of the two particles. In one spatial dimension, $g(r)$ is identical to the distribution of separations $P(|\delta x_t|=r)$ discussed in \Secref{fractal}. Our analysis of the one dimensional case shows that in an expanding system out of equilibrium, the radial distribution function is constant, for small enough spatial separations and finite times. Therefore we expect a competition between the two different scalings in the radial distribution function, that may affect the collision rate. For small times, so that $2a<\ee^{-\zeta t/2}$, the plateau in $P(\hat R_t = r)$  gives $g(2a) \sim \text{const}$ in one dimension, while for larger times  $g(2a) \sim (2a)^{\hat D_2-d}$.

For heavy particles in turbulence, there may be sub-regions of high particle concentration that temporarily expand into particle void regions, without being affected by the boundaries of the system. These are in a transient (non-steady) state, where caustics contribute to collision rates between particles not only through the rate of caustic formation $J$ and an increased collision velocity \cite{Gus14c,Pum16}, but possibly also through finite-time contributions to the radial distribution function $g(r)$.
\subsection{Wave propagation in disordered media}\seclab{waves}
Now we explore the connection between our results for the dissipative problem \eqnref{eom_dimless} and wave propagation in random media. Random focusing and spatial patterns of optical \cite{Ber77,Ber80}, acoustic \cite{Whi84,Kul82,Wol00}, and quantum-mechanical \cite{Topinka,Kap02,Met10} waves in disordered media can be understood in terms of their ray dynamics, governed by the dissipation-free limit, $\zeta \to 0$, of \Eqnref{eom_dimless}. In this limit, phase-space volumes are conserved in time so that there is no fractal clustering.  Fundamental quantities used to describe the wave patterns are the rate of caustic formation \cite{Kap02,Met10}, and the distribution of local stretching factors \cite{Kap02}, simply $|\delta x_t/\delta x_{\tg}|$ in one spatial dimension. The dissipation-free limit of \Eqnref{eom_dimless} also arises in the analysis of Anderson localisation in one-dimensional disordered quantum systems \cite{Hal65,Thouless,Sch02}, where the spatial FTLE describes fluctuations and decay of wave-function amplitudes \cite{Gus16}. 

Equation~\eqnref{hi} shows that the linear part in the spatial rate function $\hat I(s)$ appears for $s < 0$. Our discussion of the distribution of spatial separations in \Secref{fractal} then implies a linear scaling in the cumulative distribution of the stretching factor $P(|\delta x_t/\delta x_{t=0}| \leq r)\sim r \ee^{t \hat I_0}$, for $r<1$. Hence, the probability density of spatial stretching factors $P(|\delta x_t/\delta x_{t=0}| = r) = \dd{r} P(|\delta x_t/\delta x_{t=0}| \leq r)$ is finite and constant for stretching factors smaller than one. 

It has been argued that the distribution of stretching factors in the dissipation-free limit is approximately log-normal \cite{Kap02}, meaning that $\log |\delta x_t/\delta_{t=0}|$ is normally distributed. This is motivated by expanding $\hat I(s)$ around its minimum, $\hat I(s) \sim \hat I''(\hat \lambda)(s-\hat \lambda)^2/2$, hence neglecting non-Gaussian fluctuations. As we have shown, this approximation fails to describe the statistics of stretching factors $|\delta x_t/\delta x_{t=0}|$ smaller than or equal to unity.

Furthermore, the spatial SCGF that describes the moments of the stretching factor, $\hat \Lambda(k) = \lim_{t\to\infty}\frac1t\log\langle |\delta x_t/\delta x_{t=0}|^k \rangle$, is sometimes assumed to be equal to the phase-space SCGF, $\hat \Lambda(k) = \Lambda(k)$, for all $k$ \cite{Sch02}. Our result for the spatial SCGF, \Eqnref{hlam}, shows that $\hat \Lambda(k)$ diverges for $k\leq-1$ also when $\zeta = 0$.  While the assumption that $\hat \Lambda(k) = \Lambda(k)$ does take into account the non-Gaussian fluctuations of the phase-space FTLEs, it does not consider the non-Gaussian fluctuations induced by the caustics. Therefore, the assumption fails to describe the long-term behaviour of the negative moments of the stretching factor.
\subsection{Deterministic chaos}\seclab{detchaos}
Related questions are of importance in classical systems that exhibit deterministic chaos with a positive maximal Lyapunov exponent. In such systems, although $\lambda_1>0$, trajectories may nevertheless stay close together in configuration space for some time, when the local stretching factors are small \cite{Werner12}. In Ref.~\cite{Sil07} the probability of zero spatial stretching, i.e., $\hat\sigma_t = 0$ or $|\delta x_t/\delta x_{t=0}|=1$ was computed for the standard map and for a randomly kicked-rotor system. The probabilities of observing $\hat\sigma_t = 0$ or $|\delta x_t/\delta x_{t=0}|=1$ are determined by the behaviour of $\hat I(s)$ at $s=0$. Using different methods, the authors of \Cite{Sil07} found that taking into account phase-space folds leads to
\algn{\eqnlab{inequals}
	\hat I(0) < \hat \lambda\,, \quad \text{and} \quad \hat I'(0) \geq -1\,.
}
These results for $\hat \sigma_t=0$ can be explained and extended using \Eqnref{hi}. To obtain the first inequality in \Eqnref{inequals}, consider the function $-s + \hat \lambda$. Using $\hat I(\hat\lambda)=\hat I'(\hat\lambda) =0$, we first find that $\hat I(s)\leq \hat \lambda-s$ for $s\leq \hat\lambda$ and $|s-\hat \lambda|\ll 1$. Because $I'(s)\geq -1$ we conclude that $\hat I(s)\leq \hat \lambda-s$ for $s\leq \hat \lambda$, where equality only holds for $s=\hat \lambda$ (see dash-dotted lines \Figref{ischem}). The second inequality in \Eqsref{inequals} is a straightforward consequence of the more general relation $\hat I'(s) \geq -1$, which holds for all $s$. Hence, we extend \Eqsref{inequals} to
\algn{
	\hat I(s)< \hat \lambda-s\,,\quad\text{for}\ s<\hat \lambda \quad \text{and} \quad \hat I'(s) \geq -1 \,, \quad\text{for all}\ s\,.
}
The inequalities in \Eqsref{inequals} then follows by setting $s=0$. In conclusion,  our results for the spatial FTLE appear to apply also to deterministic, chaotic dynamics. This is perhaps not unexpected, as statistical descriptions are suitable for chaotic systems that are sufficiently mixing \cite{Ott02}. We conclude that the rate function $\hat I(s)$ for the Hamiltonian systems analysed in \Cite{Sil07} has a linear part starting at $s=0$ and extending to $s\to-\infty$.
\section{Conclusions}\seclab{conclusions}
In this paper we quantified the effects of fractal catastrophes (caustics that arise in the projection of a dynamical fractal attractor) upon spatial clustering of inertial particles in a random force field. For one spatial dimension, we showed that these caustics lead to an exponential increase of the probability to observe large negative spatial FTLE, resulting in a universal law of spatial clustering. 

We demonstrated that caustics give rise to a universal negative tail in the rate function for
the distribution of the spatial FTLE distribution (the rate function is essentially the logarithm of this distribution). This universal linear part of the rate function implies that the moments of spatial separations below a critical order diverge in finite time, and that the spatial correlation dimension $\hat D_2$ obeys the projection formula $\hat D_2 = \min\{D_2,1\}$.
Our theory shows how the distribution of spatial separations evolves as a function of time,
and how it approaches its steady state. Folds of the phase-space manifold and fractal clustering affect this distribution in two distinct ways: caustics cause the distribution of spatial separations to become constant
at small separations, and fractal clustering gives rise to a self-similar regime. When the spatial correlation dimension is less than the spatial dimension then these two regimes are well separated. Otherwise
the regimes overlap. 

For white-in-time Gaussian force fields we calculated the distribution of phase-space FTLEs explicitly. This distribution exhibits a fluctuation relation, associating the probabilities of phase-space regions with large positive and large negative FTLE, voids and clusters. Our exact results imply that this balance is destroyed in the distribution of the spatial FTLE, a consequence of increased clustering due to caustics. 

We showed that our results have implications for different problems in statistical physics and chaos theory, where they allow to explain and extend existing results and put into question some of the approximations.

We obtained all results that characterise spatial quantities from the spatial rate function $\hat I(s)$, which acquires a universal linear part under spatial projection. Therefore, the expressions for the spatial rate function $\hat I(s)$ and of its Legendre transform $\hat \Lambda(k)$ are the main results of this paper. 

An open question is how the conceptual insights obtained in one dimension carry over to systems in higher dimensions. A complete analysis, as provided for $d=1$ in this paper, is challenging because it involves a much more complex dynamics. However, our results should extend to the growth rate $t^{-1}\log\mc{\hat V}_t$ of an infinitesimal $d$-dimensional spatial volume $\mc{\hat V}_t$ quantifying the long-time probability of observing local particle-rich regions in configuration space. We speculate that the rate function of this growth rate has a universal linear negative tail, resulting in an exponentially increased probability of particle clusters similar to the one-dimensional case.

Important insights into how catastrophes shape the divergence of $\hat \Lambda(k)$ for $k\leq -1$ could be obtained by studying the effect of a finite cutoff upon $\hat \Lambda(k)$. 
Berry \cite{Ber77} considered the effect of a cutoff given by the wavelength of
light upon the patterns of light  intensity  $I(\ve x,t)$  focused by a random medium. 
The cutoff removes the divergencies for non-zero values of $\lambda$. Berry calculated 
how the intensity moments diverge as $\lambda\to 0$, and computed the
 critical exponents $\nu_n$ associated with the $n$-th intensity moment. He
 showed that as $n$ increases, the critical exponents are dominated by contributions from catastrophes of increasing codimensions. 
In our model the physical origin of a cutoff is different, for example due to a finite number of particles.  
We expect that such a cutoff regularises the divergence of $\hat \Lambda(k)$ for $k\leq -1$, and that
it should be possible to calculate the  corresponding critical exponents.
 They could give further insights into the divergencies caused by caustics in dissipative systems of the kind discussed here, and will yield a better understaning of the effect of fractal catastrophes on spatial clustering.

\ack
We thank Stellan \"Ostlund and Anshuman Dubey for their comments on the manuscript. JM, KG, and BM were supported by the grant {\em Bottlenecks for particle growth in turbulent aerosols} from the Knut and Alice Wallenberg Foundation, Dnr. KAW 2014.0048, and in part by VR grant no. 2017-3865. Computational resources were provided by C3SE and SNIC.

\appendix
\section{Derivation of $\hat I(s)$}\applab{hi}
We now derive a large deviation principle for $\hat \sigma_t$ with rate function $\hat I(s)$ without making explicit use of the equation of motion \eqnref{projx}. This allows us to circumvent the difficulties associated with the finite-time divergences of $\hat \sigma_t$. The large-deviation form of the cumulative distribution of the spatial FTLE $\hat \sigma_t$ reads
\algn{\eqnlab{ldp}
    P({\hat \sigma}_t \leq s) {\propto} \exp\big[-t\inf_{s'\leq s} \hat I(s')\big]\,,
}
with the spatial rate function $\hat I(s)$ yet to be determined. To find $\hat I(s)$, we first write for $P(\hat\sigma_t \leq s)$:
\algn{\eqnlab{hsigma}
	P(\hat\sigma_t \leq s) 	&= P\left(\ee^{t\hat\sigma_t} \leq \ee^{ts}\right)
						= P\left(\frac{\cos\alpha_t}{\cos\alpha_{t=0}}\ee^{t\sigma_t^{(1)}} \leq \ee^{ts}\right)
						= P\left(\cos\alpha_t \leq \exp[t(s-\sigma^{(1)}_t)]\right)\,.
}
Here we used that $|\delta { x}_t/\delta {x}_{\tg}| = \cos \alpha_t \,\ee^{t\sigma ^{(1)}_t}= \ee^{t\hat \sigma_t}$, and that $\cos\alpha_{t=0}=1$.
Now, as discussed in \Secref{ps_ftle}, $\alpha_t$ passes $-\pi/2$ with rate $J$. The large-deviation principle \eqnref{margftle}, on the other hand, implies that $\sigma^{(1)}_t$ stabilises in the vicinity of the maximal phase-space Lyapunov exponent $\lambda_1$. Therefore, the joint distribution of $\alpha_t$ and $\sigma^{(1)}_t$ factorises. As a consequence, the cumulative distribution function of $\cos\alpha_t$ conditional on a $\sigma^{(1)}_t$ simplifies to
\algn{\eqnlab{factorisation}
P\left(\cos\alpha_t \leq\exp[t(s-s')]\big| \sigma^{(1)}_t = s'\right) &\approx 	P_\text{st}\left(\cos\alpha_t \leq\exp[t(s-s')]\right)\,.
}
We now condition \Eqnref{hsigma} on $\sigma^{(1)}_t = s'$ and use \Eqnref{factorisation} to simplify the expression:
\algn{\eqnlab{cumhs}
	P(\hat\sigma_t \leq s) 	&= \int_{-\infty}^\infty\ed s' P\left(\hat \sigma_t \leq s|\sigma_t^{(1)}=s'\right)P(\sigma^{(1)}_t = s') \,\nn \\
						&= \int_{-\infty}^\infty\ed s' P\left(\cos \alpha_t \leq\exp[t(s-s')]\big| \sigma_t^{(1)}=s'\right)P(\sigma^{(1)}_t = s') \,\nn \\
						&\propto \int_{-\tfrac{\zeta}{2}}^\infty\ed s' P_\text{st}\left(\cos \alpha_t \leq\exp[t(s-s')]\right)\ee^{-tI(s')}.
}
In the second step we used \Eqnref{factorisation}, and in the last step we inserted the large-deviation form of $\sigma^{(1)}_t$, \Eqnref{margftle}. We now split the integral in \Eqnref{cumhs} into two parts: $s'<s$ and $s'>s$, so that for for large times, $\exp[t(s-s')]\gg 1$ and $\exp[t(s-s')]\ll1$, respectively. For $\exp[t(s-s')]\gg 1$ we have, trivially, $P_\text{st}(\cos\alpha_t\leq \exp[t(s-s')]) = 1$. For $\exp[t(s-s')]\ll1$, i.e. $s'>s$, we find using \Eqnref{pst}, $P_\text{st}(\cos\alpha_t\leq \exp[t(s-s')]) \sim 2J\exp[t(s-s')]$. Putting these results together we obtain
\algn{\eqnlab{pcum}
	P(\hat \sigma_t \leq s) &\propto \int_{-\zeta/2}^{s}\ed s' \exp[-t I(s')] + 2J\exp(ts)\int_s^\infty \ed s' \exp(-t[I(s')+s'] )	\nn \\
					&\propto\exp\left[-t\inf_{s'\leq s_+}I(s')\right] + 2J \exp\left[-t\left(-s +\inf_{s'\geq s_+}\{I(s')+s'	\} \right)\right]\,,
}
where we defined $s_+ = \max\{s,-\zeta/2\}$. Although we are interested only in the exponential growth rate of $P(\hat \sigma_t \leq s)$, we kept the prefactor $2J$ in \Eqnref{pcum} to show that the second term vanishes when $J=0$. For $J>0$ the relative size of the two exponents determines which of the two terms in \Eqnref{pcum} is the leading one. This, in turn, is determined by the location $s^*$ of the infimum of $I(s)+s$, 
\algn{
s^* = \text{argmin}_{s'\in \mbb{R}}\{I(s') +s'\}\,.
}
Assuming that $I(s)$ is differentiable and convex, $s^*$ is determined uniquely by the implicit equation
\algn{\eqnlab{imps0}
	I'(s^*) = -1\,.
}
Now if $s^*<s_+$, then $\inf_{s'\geq s_+}\{I(s')+s'\} = I(s_+) +s_+$. It follows that the first term in \Eqnref{pcum} is the leading one as $t\to\infty$. This implies that
\algn{
	P(\hat \sigma_t \leq s) \propto \exp\left[-t\inf_{s'\leq s_+}I(s')\right]\,.
}
If, on the other hand, $s_+\leq s^*$, then the second term is the leading one, so that
\algn{
	P(\hat \sigma_t \leq s) \propto 2J\exp\left[-t(-s -\hat I_0 )\right] = 2J\exp\left[-t\inf_{s'\leq s}\{-s' -\hat I_0 \}\right]\,.
}
The value of the constant $\hat I_0$ depends on the location of $s^*$ relative to $-\zeta/2$:
\algn{\eqnlab{result2}
	\hat I_0 = \css{	\Lambda(-1)\,,	&	s^*\geq-\zeta/2\,,	\\
			 \zeta/2- I(-\zeta/2)\,,	& s^*<-\zeta/2\,.}
}
The scaled cumulant-generating function at $k=-1$, $\Lambda(-1)$, appears in \Eqnref{result2} because by definition $\Lambda(-1) = \sup_{s\in\mbb{R}}\left\{-s-I(s)	\right\} = -\inf_{s\in\mbb{R}} \left\{I(s) +s	\right\}$. We conclude, for general $s$ and using \Eqnref{ldp},
\algn{\eqnlab{hiapp}
	\hat I(s) = \css{I(s)\,, 	& s \geq \max\{s^*,-\zeta/2\}\,,\\
						-s-\hat I_0\,,		& \text{otherwise}\,.
				}
}
This is \Eqnref{hi} in the main text.
\section{Derivation of $\hat \Lambda(k)$ by Legendre transform}\applab{hlam}
We compute $\hat \Lambda(k)$ from the Legendre transform \eqnref{leg}. When $\hat I(s)$ is differentiable in $s$, then $\hat\Lambda(k)$ is uniquely determined by inverting $k = \hat I'(s)$. This is the case for $s^*\geq-\zeta/2$ where we find using \Eqsref{hi} and \eqnref{leg}:
\algn{\eqnlab{hlamapp}
	\hat \Lambda(k) = \css{	\Lambda(k)\,, 	& k>-1\\
						\infty\,, 		& k\leq-1\,.}
}
The case $s^*<-\zeta/2$ is slightly more complicated. Since $\hat I(s)$ is not continuously differentiable at $s=-\zeta/2$, we must consider the left and right limits of the derivative at this point. Approaching $s=-\zeta/2$ from the left we find $\hat I'([\zeta/2]^-)=-1$, a direct consequence of the linear part. For the derivative approaching from the right we obtain $\hat I'([-\zeta/2]^+) = I'([-\zeta/2]^+)>-1$ and we denote the value of the right derivative by {$k^*$}. The gap between the left and the right derivative, $-1 < {k^*}$, implies that the equation $k = \hat I'(s)$ does not have a solution for $k$-values in the interval $k\in(-1,{k^*})$. Therefore, the value for $s$ which maximises the right-hand side of \Eqnref{leg} stays equal to $s=-\zeta/2$ over the interval $k\in(-1,{k^*})$. Hence, we find for $k\in(-1,{k^*})$, $\hat \Lambda(k) = -\zeta k/2 - \hat I(-\zeta/2)$, which is linear in $k$. In conclusion, we find for $s^*<-\zeta/2$:
\algn{\eqnlab{hlamapp2}
	\hat \Lambda(k) = \css{			\Lambda(k)\,,	&	k\geq {k^*}	\\
								-k\zeta/2- I(-\zeta/2)\,,	&	{k^*}>k>-1	\\
								\infty\,,					& k\leq-1\,.
					}
}
Putting \Eqsref{hlamapp} and \eqnref{hlamapp2} together in one equation we obtain \Eqsref{hlam} and \eqnref{kstar} in the main text.
\section{Calculation of $\Lambda(k)$ for even integer $k$}\applab{lamk}
We find exact expressions for the spectrum of the tilted generator $\mc{L}_k$ for even integer $k$ and show that $\Lambda(k)$ is real for even values of $k$. The tilted generator and its adjoint obey the eigenvalue equations \eqnref{tgen}, with the phase-space SCGF $\Lambda(k)$ as the leading eigenvalue. We apply to \Eqsref{tgen} the transformations
\algn{
	r_k(z) &= (z^2 + 1)^{-k/2} \tilde r_k(z)\,, \\ 	
	l_k(z) &= (z^2 + 1)^{k/2} \tilde l_k(z)\,,
}
and introduce the change of variables $z\to y = z+\zeta/2$. The functions $\tilde r_k(y)$ and $\tilde l_k(y)$ then obey the eigenvalue equations
\sbeqs{\eqnlab{op}
\algn{
	\varepsilon^2\zeta^3\tilde r_k''(y)-\left(y^2-\frac{\zeta^2}{4}\right)\tilde r_k'(y) + ky\tilde r_k(y)&= \left(	\Lambda(k) + \frac{\zeta k}{2}		\right)\tilde r_k(y)\,,	\eqnlab{op1}\\
	\varepsilon^2\zeta^3\tilde l_k''(y) + \left(y^2 - \frac{\zeta^2}4\right)\tilde l'_k(y) + (k+2)y \tilde l_k(y) &= \left(	\Lambda(k) + \frac{\zeta k}{2}		\right) \tilde l_k(y)\,. \eqnlab{op2}
}
}
Now, we rescale $y$ by $y \to \zeta y$ and follow a method described in \Cite{Sch02}. That is, we first write $\tilde r_k(y)$ as a polynomial in $y$
\algn{\eqnlab{rtilde2}
	\tilde r_k(y) = \sum_{n=0}^N a_n y^n\,,
}
where we choose $a_N = 1$. Substituting \eqnref{rtilde2} into \eqnref{op1}, we obtain a recurrence relation for $a_n$ which terminates at $N=k$, for positive integer $k$. In order to satisfy the boundary conditions that $r_k(z)$ must to be symmetric for large argument, we need to restrict $N$ to positive \textit{even} integer $k$.  The recurrence relation \eqnref{rtilde2} can be written as an eigenvalue problem for the vector $\mathbf{a} = (a_0,a_1,\ldots,a_{k-1},1)$:
\algn{
	\Lambda(k)\mathbf{a} = \mbb{L} \mathbf{a}\,,
}
with the $(k+1)\times(k+1)$-dimensional matrix $\mbb{L}$ given by the matrix:
\algn{
	\mbb{L} = \mat{
	-\frac{k}2 &	\frac{1}4	& 2\varepsilon^2 & 0 & \cdots& \cdots&\cdots&\cdots&0	\\ 
	k&-\frac{k}2&\frac{2}4&6\varepsilon^2& 0& \cdots&\cdots & \cdots& \vdots	\\
	0&k-1&\ddots&\ddots&\ddots&\ddots&\cdots&\cdots&\vdots	\\
	\vdots&0&\ddots&\ddots&\ddots&\ddots&\ddots&\ddots&0	\\
	\vdots&\vdots&\ddots&\ddots&\ddots&\ddots&\ddots&\frac{k-1}{4}&k(k-1)\varepsilon^2	\\
	0&\ddots&\cdots&\cdots&\cdots&0&2&-\frac{k}{2}&\frac{k}4	\\
	0&\cdots&\cdots&\cdots&\cdots&\cdots&0&1&-\frac{k}{2}
	}
}
The matrix $\mbb{L}$ is of the Metzler type, which means that all its off-diagonal entries are non-negative. For this kind of matrix one can prove that its largest eigenvalue (and thus $\Lambda(k)$ for positive, even integer $k$), is strictly real \cite{McC00}. Using the fluctuation relation \Eqnref{fluct21}, we can extend this result to negative even integer $k$. We conclude that $\Lambda(k)$ is real for all even integer $k$, as stated in the main text. The approach described here can be used to write $\Lambda(k)$ as the dominant root of a polynomial of order $k+1$ for all finite and even values of $k$. For $k=2$ and $k=4$ we obtain after reversing the rescaling $\Lambda(k) \to \Lambda(k)/\zeta$:
\sbeqs{\eqnlab{exeig}
\algn{
	0=&-\frac{\Lambda(2)^3}{2} -\frac{3\zeta\Lambda(2)^2 }{2}-\zeta^2\Lambda(2)+2 \varepsilon ^2 \zeta^3	\,,	\\
	0=&-\frac{\Lambda(4)^5}{24\zeta^2}-\frac{5 \Lambda(4)^4}{12\zeta}-\frac{35 \Lambda(4)^3}{24}+\zeta\Lambda(4)^2 \left(\frac{7 \varepsilon ^2}{2}-\frac{25}{12}\right)
	+\zeta^2\Lambda(4) \left(14 \varepsilon ^2-1\right)+12 \varepsilon ^2\zeta^3\,.
}
}
The largest solutions of \Eqsref{exeig} are real and yield implicit expressions for $\Lambda(2)$ and $\Lambda(4)$.
\section{Derivation of fluctuation relation \eqnref{fluct2} using the tilted generator}\applab{fluct1}
We derive the fluctuation relations \Eqnref{fluct2} from the eigenvalue equations \Eqnref{tgen}. We start from the transformed \Eqsref{op}. The idea is to bring the first equation into the same form as the second one by a suitable change of variables and then compare the corresponding largest eigenvalues $\Lambda(k)$.  To this end we transform $y\to-y$ and $k\to-k-2$ in \Eqnref{op2} to obtain
\algn{
	&\varepsilon^2\zeta^3\tilde l_{-k-2}''(-y) - \left(y^2 - \frac{\zeta^2}4\right) \tilde l'_{-k-2}(-y) + k y \tilde l_{-k-2}(-y)\nn \\
	&= \left(	\Lambda(-k-2) - \frac{\zeta k}{2}-\zeta		\right) \tilde l_{-k-2}(-y)\,.
}
Assuming non-degeneracy of the leading eigenvalue, we compare this equation to \eqnref{op1} and obtain:
\algn{\eqnlab{fluct1ap}
	\Lambda(k) + \frac{\zeta k}{2}	 = 	\Lambda(-k-2) - \frac{\zeta k}{2}-\zeta\,.		
}
The fluctuation relation [\Eqnref{fluct21} in the main text] follows directly from the shift $k\to k-1$:
\algn{\eqnlab{fluct2ap}
	\Lambda(k-1) + \zeta\frac{k-1}{2}	 &= 	\Lambda(-k-1) - \zeta\frac{k-1}{2}-\zeta \,,\nn \\
	\Lambda(k-1)- \Lambda(-k-1)&=  - \zeta k
}
 The fluctuation relation for $I(s)$, \Eqnref{fluct22} in the main text, is obtained from that of $\Lambda(k)$ by Legendre transform:
 \algn{
 	I(s-\zeta/2) 	&= \sup_{k\in\mbb{R}}\left\{ k(s-\zeta/2) - \Lambda(k)\right\}\,,\nn \\
				&= \sup_{k\in\mbb{R}}\left\{ -(k+1)(s-\zeta/2) - \Lambda(-k-1)\right\}\,,\nn \\
			&= \sup_{k\in\mbb{R}}\left\{(k-1)(-s-\zeta/2)+\zeta k-2s	-\Lambda(k-1)-\zeta k\right\} \,,\nn \\
			&= I(-s-\zeta/2) - 2s\,. \eqnlab{ifluct}
 }
\section{Connection between fluctuation relations \eqnref{fluct2} and time-reversal invariance}\applab{fluct2}
We derived the fluctuation relations \eqnref{fluct2} from the eigenvalue problem of the tilted generator $\mathscr{L}^\dagger$ in phase space. More generally it can be shown that fluctuation relations of the type \eqnref{fluct2} follow from the statistical invariance under time reversal~\cite{Fou07,Che08}. To make the connection we use the method described in \Cite{Che08} for Markov processes to show that the fluctuation relations \eqnref{fluct2} have their origin in the time-reversal invariance of the shifted process $Y_t = Z_t + \zeta/2$, which obeys the dynamics
\algn{\eqnlab{y}
	\dot Y_t = -{Y_t}^2 + \frac{\zeta^2}{4} + F_t\,.
}
If the statistics of the force gradient $F_t$ is invariant under time-reversal, the equation of motion \eqnref{y} is invariant under the transformation $(t,Y_t) \to (-t,-Y_{-t})$. Using the procedure described in \Cite{Che08} we identify the observable associated with this symmetry as
\algn{\eqnlab{w}
	\mc{W}_t 	 = 2 \int^t_0 {\ed t'}\,(Z_{{t'}} +\tfrac{\zeta}{2} ) - \Delta \varphi\,,
}
where 
\begin{equation}
\label{eq:delta_phi_def}
\Delta \varphi = \log \rho_t(Z_t+\zeta/2) - \log \rho_{\tg}(Z_{\tg}+\zeta/2)\,.
\end{equation}
Here $\rho_t$ and $\rho_{\tg}$ are the initial and final densities of $Z_t$. According to \Cite{Che08} this implies that the rate function $I_{\mc{W}_t}$ corresponding to the observable $\mc{W}_t/t$ has the symmetry
\algn{\eqnlab{wfluct}
	I_{\mc{W}_t} ({s}) -I_{\mc{W}_t}(-{s}) = -{s}\,.
}
Having identified the observable \eqnref{w}, and using the general result \eqnref{wfluct}, we can now show the fluctuation relation for $\sigma^{(1)}_t$. To this end, we use \Eqnref{sigma2} in the main text to express $\sigma^{(1)}_t$ in terms of $\mc{W}_t$ as
\algn{\eqnlab{W_t}
	\sigma^{(1)}_t&=\frac1{t} \int_0^t \ed t' Z_{t'} + \frac1t \int_0^t \ed t' \dd{t'}\log\sqrt{Z_{t'}^2 + 1}\,,\nn \\
	 &=\frac12\left(\frac{\mc{W}_t}{t} - \zeta\right) + \frac{\Delta \varphi}{t}\,.
}
To obtain the second equality, we choose $\rho_t(Z_t+\tfrac{\zeta}{2})$ and $\rho_{t=0}(Z_{t=0}+\tfrac{\zeta}{2})$ in Eq.~(\ref{eq:delta_phi_def}) as follows:
\algn{
\rho_t(Z_t+\tfrac{\zeta}{2}) =\frac{1}{\pi(Z_t^2 + 1)}\quad\mbox{and}\quad \rho_{t=0}(Z_{t=0}+\tfrac{\zeta}{2})  =\frac{1}{\pi(Z_{t=0}^2 + 1)}\,.
}
From \Eqsref{wfluct} and \eqnref{W_t} we then conclude that the rate function $I(s)$ of $\sigma^{(1)}_t$ must obey:
\algn{
	I_{\mc{W}_t}(s) = I\left({s/2 -\zeta/2}\right)\,, \quad I_{\mc{W}_t}(-s) = I\left({-s/2-\zeta/2}\right)\,,
}
so that \Eqnref{wfluct} implies the the fluctuation relation for $I(s)$, \Eqnref{fluct22} in the main text. This shows that the phase-space fluctuation relations \eqnref{fluct2} follow from statistical invariance of the Markov dynamics under time-reversal symmetry, as outlined in \Cite{Che08}.

\vspace*{5mm}

%\bibliographystyle{unsrt}
%\bibliography{ldt_refs_21}

\begin{thebibliography}{10}

\bibitem{Bodenschatz2010}
E.~Bodenschatz, S.~P. Malinowski, R.~A. Shaw, and F.~Stratmann.
\newblock Can we understand clouds without turbulence?
\newblock {\em Science}, 327(5968):970--971, 2010.

\bibitem{Wilkinson_2008}
M.~Wilkinson, B.~Mehlig, and V.~Uski.
\newblock Stokes trapping and planet formation.
\newblock {\em The Astrophysical Journal Supplement Series}, 176(2):484--496,
  2008.

\bibitem{Anders}
A.~Johansen, J.~Blum, H.~Tanaka, C.~Ormel, M.~Bizzaro, and H.~Rickman.
\newblock In H.~Beuther, R.~S. Klessen, C.~P. Dullemond, and T.~Henning,
  editors, {\em Protostars \& Planets {VI}}, University of Arizona Press, 2014.

\bibitem{Som93}
J.~C. Sommerer and E.~Ott.
\newblock Particles floating on a moving fluid: A dynamically comprehensible
  physical fractal.
\newblock {\em Science}, 259(5093):335--339, 1993.

\bibitem{Bec03}
J.~Bec.
\newblock {Fractal clustering of inertial particles in random flows}.
\newblock {\em Physics of Fluids}, 15(11):81--84, 2003.

\bibitem{Meh04}
B.~Mehlig and M.~Wilkinson.
\newblock Coagulation by random velocity fields as a {K}ramers problem.
\newblock {\em Phys. Rev. Lett.}, 92, 2004.
\newblock 250602.

\bibitem{Gus16}
K.~Gustavsson and B.~Mehlig.
\newblock {Statistical models for spatial patterns of heavy particles in
  turbulence}.
\newblock {\em Advances in Physics}, 65(1):1--57, 2016.

\bibitem{Woo05}
A.~M. Wood, W.~Hwang, and J.~K. Eaton.
\newblock Preferential concentration of particles in homogeneous and isotropic
  turbulence.
\newblock {\em Int. J. Multiphase Flow}, 31:1220--1230, 2005.

\bibitem{Saw08}
E.-W. Saw, R.A. Shaw, S.~Ayyalasomayajula, P.Y. Chuang, and A.~Gylfason.
\newblock Inertial clustering of particles in high-reynolds-number turbulence.
\newblock {\em Phys. Rev. Lett.}, 100(21):214501, 2008.

\bibitem{Sal08}
J.P.L.C Salazar, J.~De Jong, L.~Cao, S.H. Woodward, H.~Meng, and L.R. Collins.
\newblock Experimental and numerical investigation of inertial particle
  clustering in isotropic turbulence.
\newblock {\em Journal of Fluid Mechanics}, 600:245--256, 2008.

\bibitem{War09}
Z.~Warhaft.
\newblock Laboratory studies of droplets in turbulence: towards understanding
  the formation of clouds.
\newblock {\em Fluid Dyn. Res.}, 41, 2009.
\newblock 011201.

\bibitem{Bal10}
S.~Balachander and J.~K. Eaton.
\newblock Turbulent dispersed multiphase flow.
\newblock {\em Ann. Rev. Fluid Mech.}, 42:111--133, 2010.

\bibitem{Mon10}
R.~Monchaux, M.~Bourgoin, and A.~Cartellier.
\newblock Preferential concentration of heavy particles: A {V}oronoi analysis.
\newblock {\em Phys. Fluids}, 22, 2010.
\newblock 103304.

\bibitem{Gib12}
M.~Gibert, H.~Xu, and E.~Bodenschatz.
\newblock Where do small weakly inertial particles go in a turbulent flow?
\newblock {\em J. Fluid Mech.}, 698:160--167, 2012.

\bibitem{Saw12b}
E.-W. Saw, R.~A. Shaw, J.~P. L.~C. Salazar, and L.~R. Collins.
\newblock Spatial clustering of polydisperse inertial particles in turbulence:
  Ii. comparing simulation with experiment.
\newblock {\em New J. Phys.}, 14, 2012.
\newblock 105031.

\bibitem{Wan93}
L.~Wang and M.~R. Maxey.
\newblock Settling velocity and concentration distribution of heavy particles
  in homogeneous isotropic turbulence.
\newblock {\em J. Fluid Mech.}, 256:27--68, 1993.

\bibitem{Hog01}
R.~C. Hogan and J.~N. Cuzzi.
\newblock Stokes and {R}eynolds number dependence of preferential particle
  concentration in simulated 3d turbulence;.
\newblock {\em Phys. Fluids}, 13:2938--2945, 2001.

\bibitem{Chu05}
J.~Chun, D.~L. Koch, S.~L. Rani, A.~Ahluwalia, and L.~R. Collins.
\newblock Clustering of aerosol particles in isotropic turbulence.
\newblock {\em J. Fluid Mech.}, 536:219--251, 2005.

\bibitem{Pic05}
M.~Picciotto, C.~Marchioli, and A.~Soldati.
\newblock Characterization of near-wall accumulation regions for inertial
  particles in turbulent boundary layers.
\newblock {\em Phys. Fluids}, 17, 2005.
\newblock 098101.

\bibitem{Sim06}
K.~Gawedzki and M.~Vergassola.
\newblock Connection between two statistical approaches for the modelling of
  particle velocity and concentration distributions in turbulent flow: The
  mesoscopic {E}ulerian formalism and the two-point probability density
  function method.
\newblock {\em Phys. Fluids}, 18, 2006.
\newblock 125107.

\bibitem{Bec06}
J.~Bec, L.~Biferale, G.~Boffetta, M.~Cencini, S.~Musacchio, and F.~Toschi.
\newblock Lyapunov exponents of heavy particles in turbulence.
\newblock {\em Phys. Fluids}, 18, 2006.
\newblock 091702.

\bibitem{Bec07}
J.~Bec, L.~Biferale, M.~Cencini, A.~Lanotte, S.~Musacchio, and F.~Toschi.
\newblock Heavy particle concentration in turbulence at dissipative and
  inertial scales.
\newblock {\em Phys. Rev. Lett.}, 98, 2007.
\newblock 084502.

\bibitem{Cal08}
E.~Calzavarini, M.~Cencini, D.~Lohse, and F.~Toschi.
\newblock Quantifying turbulence-induced segregation of inertial particles.
\newblock {\em Phys. Rev. Lett.}, 101, 2008.
\newblock 084504.

\bibitem{Cal08b}
E.~Calzavarini, M.~Kerscher, D.~Lohse, and F.~Toschi.
\newblock Dimensionality and morphology of particle and bubble clusters in
  turbulent flow.
\newblock {\em J. Fluid Mech.}, 607:13--24, 2008.

\bibitem{Col09}
S.~W. Coleman and J.~C. Vassilicos.
\newblock A unified sweep-stick mechanism to explain particle clustering in
  two- and three-dimensional homogeneous, isotropic turbulence.
\newblock {\em Phys. Fluids}, 21, 2009.
\newblock 113301.

\bibitem{Saw12a}
E-W. Saw, J.~P. L.~C. Salazar, L.~R. Collins, and R.~A. Shaw.
\newblock Spatial clustering of polydisperse inertial particles in turbulence:
  I. comparing simulation with theory.
\newblock {\em New J. Phys.}, 14, 2012.
\newblock 105030.

\bibitem{Rea00}
W.~C. Reade and L.~R. Collins.
\newblock Effect of preferential concentration on turbulent collision rates.
\newblock {\em Phys. Fluids}, 12:2530--2540, 2000.

\bibitem{And07}
B.~Andersson, K.~Gustavsson, B.~Mehlig, and M.~Wilkinson.
\newblock Advective collisions.
\newblock {\em Europhys. Lett.}, 80:69001, 2007.

\bibitem{Bec14}
J.~Bec, H.~Homann, and G.~Krstulovic.
\newblock Clustering, fronts, and heat transfer in turbulent suspensions of
  heavy particles.
\newblock {\em Phys. Rev. Lett.}, 112:234503, 2014.

\bibitem{Kru17}
J.~Kr{\"u}ger, N.E.L. Haugen, D.~Mitra, and T.~L{\o}v{\aa}s.
\newblock The effect of turbulent clustering on particle reactivity.
\newblock {\em Proceedings of the Combustion Institute}, 36(2):2333--2340,
  2017.

\bibitem{Ren70}
A.~Renyi.
\newblock {\em Probability Theory}.
\newblock Kiado, Budapest, 1970.

\bibitem{Gra83}
P.~Grassberger.
\newblock Generalized dimensions of strange attractors.
\newblock {\em Physics Letters A}, 97(6):227--230, 1983.

\bibitem{Hen83}
H.~G.~E. Hentschel and I.~Procaccia.
\newblock The infinite number of generalized dimensions of fractals and strange
  attractors.
\newblock {\em Physica D}, 8(3):435--444, 1983.

\bibitem{Bec04}
J.~Bec, K.~Gawedzki, and P.~Horvai.
\newblock {Multifractal clustering in compressible flows}.
\newblock {\em Phys. Rev. Lett.}, 92(22):224501, 2004.

\bibitem{Gra86}
P.~Grassberger.
\newblock Estimating the fractal dimensions and entropies of strange
  attractors.
\newblock {\em Chaos}, 1:291--311, 1986.

\bibitem{Bad87}
R.~Badii and A.~Politi.
\newblock Renyi dimensions from local expansion rates.
\newblock {\em Phys. Rev. A}, 35(3):1288, 1987.

\bibitem{Ell07}
R.~S. Ellis.
\newblock {\em {Entropy, large deviations, and statistical mechanics}}.
\newblock Springer, 2007.

\bibitem{Hol08}
F.~{Den Hollander}.
\newblock {\em {Large deviations}}, volume~14.
\newblock American Mathematical Soc., 2008.

\bibitem{Tou09}
H.~Touchette.
\newblock {The large deviation approach to statistical mechanics}.
\newblock {\em Physics Reports}, 478(1):1--69, 2009.

\bibitem{Ekdahl}
C.~Ekdahl.
\newblock Using large-deviation theory to estimate the correlation dimension of
  small, heavy particles in random flows, {MS}c thesis ({U}niversity of
  {G}othenburg), 2017.

\bibitem{Bal99}
E.~Balkovsky and A.~Fouxon.
\newblock {Universal long-time properties of Lagrangian statistics in the
  Batchelor regime and their application to the passive scalar problem}.
\newblock {\em Phys. Rev. E}, 60:4164--4174, 1999.

\bibitem{Fal01}
G.~Falkovich, K.~Gawedzki, and M.~Vergassola.
\newblock {Particles and fields in fluid turbulence}.
\newblock {\em Reviews of Modern Physics}, 73(4):913--975, nov 2001.

\bibitem{Bal01}
E.~Balkovsky, G.~Falkovich, and A.~Fouxon.
\newblock {Intermittent distribution of inertial particles in turbulent flows}.
\newblock {\em Phys. Rev. Lett.}, 86:2790--2793, 2001.

\bibitem{Fal02}
G.~Falkovich, A.~Fouxon, and M.~Stepanov.
\newblock {Acceleration of rain initiation by cloud turbulence}.
\newblock {\em Nature}, 419:151, 2002.

\bibitem{Wil03}
M.~Wilkinson and B.~Mehlig.
\newblock {Path coalescence transition and its applications}.
\newblock {\em Phys. Rev. E}, 68:040101, 2003.

\bibitem{Wil05}
M.~Wilkinson and B.~Mehlig.
\newblock {Caustics in turbulent aerosols}.
\newblock {\em Europhys. Lett.}, 71(2):186--192, 2005.

\bibitem{Wil06}
M.~Wilkinson, B.~Mehlig, and V.~Bezuglyy.
\newblock Caustic activation of rain showers.
\newblock {\em Phys. Rev. Lett.}, 97(4):048501, 2006.

\bibitem{Gus11b}
K.~Gustavsson and B.~Mehlig.
\newblock {Distribution of relative velocities in turbulent aerosols}.
\newblock {\em Phys. Rev. E}, 84:045304, 2011.

\bibitem{Ber77}
M.~V. Berry.
\newblock Focusing and twinkling: critical exponents from catastrophes in
  non-gaussian random short waves.
\newblock {\em Journal of Physics A: Mathematical and General}, 10(12):2061,
  1977.

\bibitem{Ber80}
M~V Berry and C~Upstill.
\newblock Iv catastrophe optics: morphologies of caustics and their diffraction
  patterns.
\newblock In {\em Progress in optics}, volume~18, pages 257--346. Elsevier,
  1980.

\bibitem{Mei17}
J.~Meibohm, L.~Pistone, K.~Gustavsson, and B.~Mehlig.
\newblock {Relative velocities in bidisperse turbulent suspensions}.
\newblock {\em Phys. Rev. E}, 96(6):061102, 2017.

\bibitem{Dun05}
K.~Duncan, B.~Mehlig, S.~{\"O}stlund, and M.~Wilkinson.
\newblock Clustering in mixing flows.
\newblock {\em Phys. Rev. Lett.}, 95(24):240602, 2005.

\bibitem{Bec08}
J.~Bec, M.~Cencini, M.~Hillerbrand, and K.~Turitsyn.
\newblock Stochastic suspensions of heavy particles.
\newblock {\em Physica D}, 237:2037--2050, 2008.

\bibitem{Wil10b}
M.~Wilkinson, B.~Mehlig, and K.~Gustavsson.
\newblock {Correlation dimension of inertial particles in random flows}.
\newblock {\em Europhys. Lett.}, 89(5):50002, 2010.

\bibitem{Gus15}
K.~Gustavsson, B.~Mehlig, and M.~Wilkinson.
\newblock {Analysis of the correlation dimension of inertial particles}.
\newblock {\em Phys. Fluids}, 27(7):073305, 2015.

\bibitem{Che08}
R.~Chetrite and K.~Gawedzki.
\newblock {Fluctuation relations for diffusion processes}.
\newblock {\em Communications in Mathematical Physics}, 282(2):469--518, 2008.

\bibitem{Che08b}
R.~Chetrite, G.~Falkovich, and K.~Gawedzki.
\newblock {Fluctuation relations in simple examples of non-equilibrium steady
  states}.
\newblock {\em Journal of Statistical Mechanics: Theory and Experiment},
  2008(08):P08005, 2008.

\bibitem{Sei12}
Udo Seifert.
\newblock {Stochastic thermodynamics, fluctuation theorems and molecular
  machines}.
\newblock {\em Reports on Progress in Physics}, 75(12):126001, 2012.

\bibitem{Cel12}
A.~Celani, S.~Bo, R.~Eichhorn, and E.~Aurell.
\newblock Anomalous thermodynamics at the microscale.
\newblock {\em Phys. Rev. Lett.}, 109:260603, 2012.

\bibitem{Topinka}
M.~A. Topinka, B.~J. LeRoy, R.~M. Westervelt, S.~E.~J. Shaw, R.~Fleischmann,
  E.~J. Heller, K.~D. Maranwoski, and A.~C. Gossard.
\newblock Coherent branched flow in a two-dimensional electron gas.
\newblock {\em Nature}, 410:183, 2001.

\bibitem{Kap02}
L.~Kaplan.
\newblock Statistics of branched flow in a weak correlated random potential.
\newblock {\em Physical Review Letters}, 89(18):184103, 2002.

\bibitem{Met10}
J.~J. Metzger, R.~Fleischmann, and T.~Geisel.
\newblock Universal statistics of branched flows.
\newblock {\em Physical {R}eview Letters}, 105(2):020601, 2010.

\bibitem{Whi84}
B.~S. White.
\newblock The stochastic caustic.
\newblock {\em SIAM Journal on Applied Mathematics}, 44(1):127--149, 1984.

\bibitem{Kul82}
V.~A. Kulkarny and B.~S. White.
\newblock Focusing of waves in turbulent inhomogeneous media.
\newblock {\em The Physics of Fluids}, 25(10):1770--1784, 1982.

\bibitem{Wol00}
M.~A. Wolfson and F.~D. Tappert.
\newblock Study of horizontal multipaths and ray chaos due to ocean mesoscale
  structure.
\newblock {\em The Journal of the Acoustical Society of America},
  107(1):154--162, 2000.

\bibitem{Sil07}
P.~G. Silvestrov and I.~V. Ponomarev.
\newblock Chaos beyond linearized stability analysis: Folding of the phase
  space and distribution of {L}yapunov exponents.
\newblock {\em Physics Letters A}, 365(4):290--294, 2007.

\bibitem{Sil06}
P.~Silvestrov.
\newblock Spectrum of the andreev billiard and giant fluctuations of the
  ehrenfest time.
\newblock {\em Phys. Rev. Lett.}, 97:067004, 2006.

\bibitem{Ose68}
V.~I. Oseledets.
\newblock A multiplicative ergodic theorem. {C}haracteristic {L}japunov,
  exponents of dynamical systems.
\newblock {\em Trudy Moskovskogo Matematicheskogo Obshchestva}, 19:179--210,
  1968.

\bibitem{Che15}
R.~Chetrite and H.~Touchette.
\newblock {Nonequilibrium Markov processes conditioned on large deviations}.
\newblock In {\em Annales Henri Poincar{\'{e}}}, volume~16, pages 2005--2057.
  Springer, 2015.

\bibitem{Tou18}
H.~Touchette.
\newblock {Introduction to dynamical large deviations of Markov processes}.
\newblock {\em Physica A}, 504:5--19, 2018.

\bibitem{Gar77}
J.~G{\"{a}}rtner.
\newblock {On Large Deviations from the Invariant Measure}.
\newblock {\em Theory of Probability {\&} Its Applications}, 22(1):24--39,
  1977.

\bibitem{Zee79}
Erik~Christopher Zeeman.
\newblock Catastrophe theory.
\newblock In {\em Structural Stability in Physics}, pages 12--22. Springer,
  1979.

\bibitem{Arn03}
Vladimir~I Arnol'd.
\newblock {\em Catastrophe theory}.
\newblock Springer Science \& Business Media, 2003.

\bibitem{Pos14}
Tim Poston and Ian Stewart.
\newblock {\em Catastrophe theory and its applications}.
\newblock Courier Corporation, 2014.

\bibitem{Kap79}
JL~Kaplan and JA~Yorke.
\newblock Functional differential equations and approximation of fixed points.
\newblock {\em Lecture notes in mathematics}, 730:204--227, 1979.

\bibitem{Ott02}
Edward Ott.
\newblock {\em Chaos in Dynamical Systems}.
\newblock Cambridge University Press, 2002.

\bibitem{Bec05}
J.~Bec.
\newblock Multifractal concentrations of inertial particles in smooth random
  flows.
\newblock {\em Journal of Fluid Mechanics}, 528:255--277, 2005.

\bibitem{Mar54}
J.~M. Marstrand.
\newblock Some fundamental geometrical properties of plane sets of fractional
  dimensions.
\newblock {\em Proceedings of the London Mathematical Society},
  s3-4(1):257--302, 1954.

\bibitem{Kau68}
Robert Kaufman.
\newblock On hausdorff dimension of projections.
\newblock {\em Mathematika}, 15(2):153--155, 1968.

\bibitem{Hun97}
Brian~R Hunt and Vadim~Yu Kaloshin.
\newblock How projections affect the dimension spectrum of fractal measures.
\newblock {\em Nonlinearity}, 10(5):1031, 1997.

\bibitem{Cri88}
Andrea Crisanti, Giovanni Paladin, and Angelo Vulpiani.
\newblock Generalized lyapunov exponents in high-dimensional chaotic dynamics
  and products of large random matrices.
\newblock {\em Journal of Statistical Physics}, 53(3):583--601, Nov 1988.

\bibitem{Pik92}
A.~S. Pikovsky.
\newblock {Statistics of trajectory separation in noisy dynamical systems}.
\newblock {\em Physics Letters A}, 165(1):33--36, 1992.

\bibitem{noteEkdahl}
C. Ekdahl explored this equation using numerical simulations of a
  one-dimensional model \cite{Ekdahl}.

\bibitem{Gus14c}
K.~Gustavsson and B.~Mehlig.
\newblock {Relative velocities of inertial particles in turbulent aerosols}.
\newblock {\em Journal of Turbulence}, 15(1):34--69, 2014.

\bibitem{Sch02}
H.~Schomerus and M.~Titov.
\newblock {Statistics of finite-time Lyapunov exponents in a random
  time-dependent potential}.
\newblock {\em Phys. Rev. E}, 66(6):066207, 2002.

\bibitem{Der07}
S.~A. Derevyanko, G.~Falkovich, K.~Turitsyn, and S.~Turitsyn.
\newblock Lagrangian and {E}ulerian descriptions of inertial particles in
  random flows.
\newblock {\em Journal of Turbulence}, 8:N16, 2007.

\bibitem{Fou07}
I.~Fouxon and P.~Horvai.
\newblock Fluctuation relation and pairing rule for {L}yapunov exponents of
  inertial particles in turbulence.
\newblock {\em Journal of Statistical Mechanics: Theory and Experiment},
  2007(08):L08002, 2007.

\bibitem{Bax88}
P.H. Baxendale and D.W. Stroock.
\newblock Large deviations and stochastic flows of diffeomorphisms.
\newblock {\em Probability Theory and Related Fields}, 80(2):169--215, 1988.

\bibitem{Hub18}
G.~Huber, M.~Pradas, A.~Pumir, and M.~Wilkinson.
\newblock {Persistent stability of a chaotic system}.
\newblock {\em Physica A}, 492:517--523, 2018.

\bibitem{Pum16}
Alain Pumir and Michael Wilkinson.
\newblock Collisional aggregation due to turbulence.
\newblock {\em Annual Review of Condensed Matter Physics}, 7:141--170, 2016.

\bibitem{Sun97}
S.~Sundaram and L.~R. Collins.
\newblock Collision statistics in an isotropic particle-laden turbulent
  suspension.
\newblock {\em J. Fluid. Mech.}, 335:75--109, 1997.

\bibitem{Hal65}
B.~I. Halperin.
\newblock Green's functions for a particle in a one-dimensional random
  potential.
\newblock {\em Phys. Rev.}, 139:A104--A117, 1965.

\bibitem{Thouless}
D.~J. Thouless.
\newblock A relation between the density of states and range of localization
  for one dimensional random systems.
\newblock {\em J. Phys. C}, 5:77--81, 1972.

\bibitem{Werner12}
M.~Wilkinson, B.~Mehlig, K.~Gustavsson, and E.~Werner.
\newblock Clustering of exponentially separating trajectories.
\newblock {\em The European Physical Journal B}, 85, 2012.

\bibitem{McC00}
C.R. MacCluer.
\newblock The many proofs and applications of {P}erron's theorem.
\newblock {\em SIAM Review}, 42(3):487--498, 2000.

\end{thebibliography}

\end{document}